\documentclass[%
 reprint,
 superscriptaddress,
 showpacs,preprintnumbers,
 amsmath,amssymb,
 pre,
]{revtex4-1}

\usepackage{graphicx}
\usepackage{dcolumn}
\usepackage{bm}
\usepackage[colorlinks,allcolors=blue]{hyperref}
\usepackage{color}
\usepackage{txfonts}

\begin{document}

\preprint{}

\title{Mode-decomposition based on crystallographic symmetry in the band-unfolding method}
 
\author{Yuji Ikeda}
\email{ikeda.yuji.6m@kyoto-u.ac.jp}
\affiliation{Center for Elements Strategy Initiative for Structure Materials (ESISM), Kyoto University, Kyoto 606-8501, Japan}

\author{Abel Carreras}
\affiliation{Department of Materials Science and Engineering, Kyoto University, Kyoto 606-8501, Japan}

\author{Atsuto Seko}
\affiliation{Center for Elements Strategy Initiative for Structure Materials (ESISM), Kyoto University, Kyoto 606-8501, Japan}
\affiliation{Department of Materials Science and Engineering, Kyoto University, Kyoto 606-8501, Japan}
\affiliation{Center for Materials Research by Information Integration, National Institute for Materials Science (NIMS), Tsukuba 305-0047, Japan}

\author{Atsushi Togo}
\affiliation{Center for Elements Strategy Initiative for Structure Materials (ESISM), Kyoto University, Kyoto 606-8501, Japan}
\affiliation{Center for Materials Research by Information Integration, National Institute for Materials Science (NIMS), Tsukuba 305-0047, Japan}

\author{Isao Tanaka}
\affiliation{Center for Elements Strategy Initiative for Structure Materials (ESISM), Kyoto University, Kyoto 606-8501, Japan}
\affiliation{Department of Materials Science and Engineering, Kyoto University, Kyoto 606-8501, Japan}
\affiliation{Center for Materials Research by Information Integration, National Institute for Materials Science (NIMS), Tsukuba 305-0047, Japan}
\affiliation{Nanostructures Research Laboratory, Japan Fine Ceramics Center, Nagoya 456-8587, Japan}

\date{\today}

\begin{abstract}
The band-unfolding method is widely used to calculate the \textit{effective} band structures 
of a disordered system from its supercell model.
The unfolded band structures show the crystallographic symmetry of the underlying structure,
where the difference of chemical components and the local atomic relaxation are ignored.
It has been, however, still difficult to decompose the unfolded band structures 
into the modes based on the crystallographic symmetry of the underlying structure,
and therefore detailed analyses of the unfolded band structures have been restricted.
In this study, we develop a procedure to decompose the unfolded band structures 
according to the small representations (SRs) of the little groups.
For this purpose, we derive the projection operators for SRs 
based on the group representation theory.
We also introduce another type of projection operators for chemical elements,
which enables us to decompose the unfolded band structures into the contributions of different combinations of chemical elements.
Using the current method, we investigate the phonon band structure of disordered face-centered cubic Cu$_{0.75}$Au$_{0.25}$,
which has large variations of atomic masses and force constants among the atomic sites due to the chemical disorder.
We find several peculiar behaviors such as discontinuous and split branches in the unfolded phonon band structure.
They are found to occur because different combinations of the chemical elements contribute to different regions of frequency.
\end{abstract}


\maketitle


\section{INTRODUCTION}
\label{sec:introduction}

Configurational disorder is commonly seen in alloy systems,
which often changes their physical properties.
First-principles calculations for such a disordered system require some approximations.
Among such approximations,
the virtual crystal approximation \cite{Bellaiche2000Virtual} and 
the coherent potential approximation \cite{Jaros1985} consider 
an effective medium for the disordered system and have often been adopted.
These methods, however,
generally do not explicitly consider the local environment around each atom,
which is sometimes critical for quantitative evaluations of physical properties of the disordered system
\cite{Lu1991Large}. 
In contrast,
the use of a supercell model to mimic the disordered system is computationally more demanding
but can accurately account for the local environment around each atom,
including local relaxation of atomic positions.
With the development of high performance computers,
the supercell approach is increasingly more popular.

A disordered system can be associated with its corresponding underlying structure,
where the difference of chemical components and the local atomic relaxation are ignored.
However, when the disordered system is mimicked by a supercell model,
it generally lacks the crystallographic symmetry of the underlying structure.
This makes it difficult to compare the band structures (of electrons and phonons) 
calculated from the supercell model with experimental data,
because they are typically described as if the disordered system has 
the crystallographic symmetry of its underlying structure.
To fill the gap between such supercell calculations and the experimental data,
a computational method to obtain the \textit{effective} band structures,
which show the symmetry of the corresponding underlying structure,
should be useful.

The band-unfolding method
\cite{Boykin2007, Ku2010, Popescu2012, Allen2013, *Allen2013E} 
is one of the methods to obtain such an effective band structure
using a supercell model.
In this method we obtain the effective band structure by decomposing 
the eigenvectors (of electrons or phonons) obtained from the supercell model 
of a disordered system 
according to the translational symmetry of its underlying crystal structure.
Already the band-unfolding method has been applied to the electronic band structures
\cite{Boykin2007,Ku2010,Popescu2012,Lee2013Unfolding,Tomic2014,Huang2014,Rubel2014,Gordienko2016}
and to the phonon band structures
\cite{Allen2013, *Allen2013E, Boykin2014, Huang2015, Zheng2016CMS, Overy2016}
of various systems with disorders
and also has been used in characteristic ways such as 
analyses of surface states
\cite{Allen2013, *Allen2013E}
and of spinor wave functions
\cite{Medeiros2015}.
Among various band-unfolding procedures,
\citeauthor{Allen2013}
use the projection operators
that decompose the eigenvectors obtained from a supercell model
according to the \textit{translational} symmetry of the underlying structure
\cite{Allen2013, *Allen2013E}.
Unlike previous approaches
\cite{Boykin2007, Ku2010, Popescu2012},
their formalization has conceptual and practical advantages
because it can be understood based on the group theory
and because it does not require any reference vectors for the decomposition.
Even in their formalization, however,
one still cannot further decompose the unfolded band structures
into the modes that transform in different ways under the symmetry operations 
with nontrivial \textit{rotational} parts.

From a group-theoretical viewpoint,
the small representations (SRs) of the little group of the wave vector $\mathbf{k}$
describe how the eigenvectors of a crystalline material at $\mathbf{k}$ transform under the symmetry operations
\cite{Kim1999, Dresselhaus2007, ElBatanouny2008}.
Since the SRs of the eigenvectors are useful to analyze various physical
behaviors such as selection rules and avoided band crossings
\cite{ITD_2013} 
in band structures,
it is reasonable to decompose the unfolded band structures 
into the modes corresponding to different SRs.
In a previous band-unfolding approach 
\cite{Boykin2014},
an unfolded band structure is decomposed according to the cumulative spectral function.
The modes determined in this manner, however, do not reflect the crystallographic symmetry 
of the underlying structure,
i.e., they do not follow any SRs of the little groups for the underlying structure.
Moreover,
the degeneracies of the modes cannot be determined in this approach 
in terms of crystallographic symmetry.
To the best of our knowledge, there have been no reports for how to decompose 
the unfolded band structures according to the SRs in a systematic manner.

In this study,
we develop a procedure to decompose unfolded band structures according to the SRs 
of the little groups of the underlying crystal structure.
For this purpose,
we introduce the projection operators for SRs [Eq.~\eqref{eq:po_srs}]
based on the group representation theory.
These projection operators can be applied to symmetrically degenerated modes.
Using the projection operators for SRs,
we can analyze the unfolded band structures in very similar ways to 
those of ordered systems in terms of crystallographic symmetry.

Here we also analyze the contributions of different chemical elements in the unfolded band structures.
In previous approaches
\cite{Boykin2007, Popescu2012, Boykin2014},
it was not clear how to investigate the contributions of different chemical elements to the unfolded band structures.
This issue largely limits our understanding about the band structures of disordered systems.
In this study,
we also derive the projection operators for chemical elements
[Eq.~\eqref{eq:pos_chemical_elements}] particularly for the phonon modes,
which enable us to decompose the unfolded phonon band structures 
according to the contribution of the chemical elements.
This decomposition enables us to analyze peculiar behaviors in the unfolded phonon band structures that are not found in ordered systems.

We use the current band-unfolding method to investigate the phonon band structure of 
disordered face-centered cubic (fcc) Cu$_{0.75}$Au$_{0.25}$.
This alloy is known to have large variations of atomic masses and force constants 
among the atomic sites due to the chemical disorder,
and hence its phonon band structure has been investigated in experimental 
\cite{Katano1988} and computational 
\cite{Dutta2010}
approaches to reveal the impacts of these variations.
We find several peculiar behaviors such as discontinuous and split branches 
in the unfolded phonon band structure.
The origins of these peculiar behaviors are analyzed based on the projection operators for SRs and those for chemical elements, both of which are derived in this study.
It is found that these peculiarities occur because different combinations of the chemical elements contribute to different regions of frequency.

\section{METHODS}
\label{sec:methods}

\newcommand{\lcg}{\bar{\mathcal{G}}^{\mathbf{k}}}

In this section,
we first summarize the computational procedure of phonon modes
because in this paper we focus on the unfolding for phonon band structures.
Next we introduce the notations to describe the relations between a supercell
model and its underlying crystal structure.
Then we derive three types of projection operators,
the keys of the current band-unfolding method.
Finally we obtain the spectral functions,
which are plotted as the unfolded band structures.

We denote transformation operators on real space points in the Seitz notation as 
$\{ \mathbf{R} | \mathbf{w} \}$,
where the rotational part $\mathbf{R}$ and the translational part $\mathbf{w}$ are 
$3 \times 3$ and $3 \times 1$ real matrices,
respectively.
$\{ \mathbf{R} | \mathbf{w} \}$ transforms a real space point $\mathbf{x}$ as 
\begin{align}
\{ \mathbf{R} | \mathbf{w} \}
\mathbf{x}
&=
\mathbf{R} \mathbf{x} + \mathbf{w}
.
\end{align}
We use the same notation also for transformation operators on functions of $\mathbf{x}$
(see Appendix~\ref{sec:transformation_of_functions} for
how $\{ \mathbf{R} | \mathbf{w} \}$ works on the functions).

\subsection{Phonon-mode calculations}
\label{sec:phonon_mode_calculations}


Suppose $\mathbf{x}_{l\kappa}$ is the equilibrium positions of the $\kappa$th
atom in the $l$th unit cell of a crystalline system.
The second-order force constants for the pair of the atoms $l\kappa$ and
$l'\kappa'$ are denoted as
$\Phi_{\alpha \beta} (l \kappa, l' \kappa')$,
where $\alpha$ and $\beta$ are indices for Cartesian coordinates.
The dynamical matrix 
$\mathbf{D} (\mathbf{K})$
at the wave vector $\mathbf{K}$
is then calculated as
\begin{align}
    D_{\kappa \kappa'}^{\alpha \beta} (\mathbf{K})
    &=
    \frac{1}{\sqrt{m_{\kappa} m_{\kappa'}}}
    \sum_{l'}
    \Phi_{\alpha \beta} (0 \kappa , l'\kappa')
    e^{i \mathbf{K} \cdot (\mathbf{x}_{l'\kappa'} - \mathbf{x}_{0 \kappa})},
\end{align}
where
$m_{\kappa}$ is the mass of the $\kappa$th atom.
Phonon frequencies 
$\omega (\mathbf{K}, J)$
and mode eigenvectors 
$\mathbf{v} (\mathbf{K}, J)$
at $\mathbf{K}$ are obtained
by solving the eigenvalue problem of 
$\mathbf{D} (\mathbf{K})$ as
\begin{align}
\mathbf{D} (\mathbf{K})
\mathbf{v} (\mathbf{K}, J)
=
[\omega (\mathbf{K}, J)]^2
\mathbf{v} (\mathbf{K}, J)
\end{align}
or
\begin{align}
\sum_{\beta \kappa'}
D_{\kappa \kappa'}^{\alpha \beta}
v_{\kappa'}^{\beta} (\mathbf{K}, J)
=
[\omega (\mathbf{K}, J)]^2
v_{\kappa}^{\alpha} (\mathbf{K}, J)
,
\end{align}
where $J$ is the band index.
$\mathbf{v} (\mathbf{K}, J)$ can be explicitly written as
\begin{align}
\mathbf{v}(\mathbf{K}, J)
=
\begin{pmatrix}
\mathbf{v}_{1}(\mathbf{K}, J) \\
\vdots \\
\mathbf{v}_{n}(\mathbf{K}, J) \\
\end{pmatrix}
=
\begin{pmatrix}
v_{1}^{x}(\mathbf{K}, J) \\
v_{1}^{y}(\mathbf{K}, J) \\
v_{1}^{z}(\mathbf{K}, J) \\
\vdots \\
v_{n}^{x}(\mathbf{K}, J) \\
v_{n}^{y}(\mathbf{K}, J) \\
v_{n}^{z}(\mathbf{K}, J) \\
\end{pmatrix}
,
\end{align}
where $n$ is the number of atoms in a unit cell,
and $3 \times 1$ matrices $\mathbf{v}_{\kappa}$ are the component for the $\kappa$th atom.
Hereafter $\mathbf{v} (\mathbf{K}, J)$ is supposed to be normalized.

To explicitly describe the dependence of the atomic displacements on 
the wave vector $\mathbf{K}$,
we consider the ``phase-weighted'' mode eigenvectors
$\tilde{\mathbf{v}} (\mathbf{K}, J)$.
The components of $\tilde{\mathbf{v}} (\mathbf{K}, J)$ is given 
for all the atomic sites in all the unit cells as
\begin{gather}
\tilde{\mathbf{v}}_{l \kappa}
(\mathbf{K}, J)
\equiv
    e^{i \mathbf{K} \cdot \mathbf{x}_{l \kappa}}
\mathbf{v}_{\kappa}
(\mathbf{K}, J)
,
\\
\tilde{\mathbf{v}}_{l}
(\mathbf{K}, J)
=
\begin{pmatrix}
\tilde{\mathbf{v}}_{l 1} (\mathbf{K}, J) \\
\vdots \\
\tilde{\mathbf{v}}_{l n} (\mathbf{K}, J)
\end{pmatrix}
.
\end{gather}
$\tilde{\mathbf{v}} (\mathbf{K}, J)$
can be regarded as 
a vector-field function of $\mathbf{x}$ defined on
$\mathbf{x}_{l \kappa}$,
and hence
$\{ \mathbf{R} | \mathbf{w} \}$
transforms
$\tilde{\mathbf{v}} (\mathbf{K}, J)$
according to Eq.~\eqref{eq:tf_vff} as
\begin{align}
    [
    \{ \mathbf{R} | \mathbf{w} \}
    \tilde{\mathbf{v}}
    (\mathbf{K}, J)
    ]_{l \kappa}
    &=
    \mathbf{R}
    \tilde{\mathbf{v}}_{l' \kappa'}
    (\mathbf{K}, J)
    ,
\end{align}
where $l'$ and $\kappa'$ satisfy
\begin{align}
    \mathbf{x}_{l' \kappa'}
    &=
    \{ \mathbf{R} | \mathbf{w} \}^{-1} 
    \mathbf{x}_{l \kappa}
    \notag
    \\
    &=
    \mathbf{R}^{-1}
    \mathbf{x}_{l \kappa}
    -
    \mathbf{R}^{-1}
    \mathbf{w}
    .
\end{align}

\subsection{Supercell model and its underlying crystal structure}

\begin{figure}[tbp]
\centering
    \includegraphics[width=0.75\linewidth]{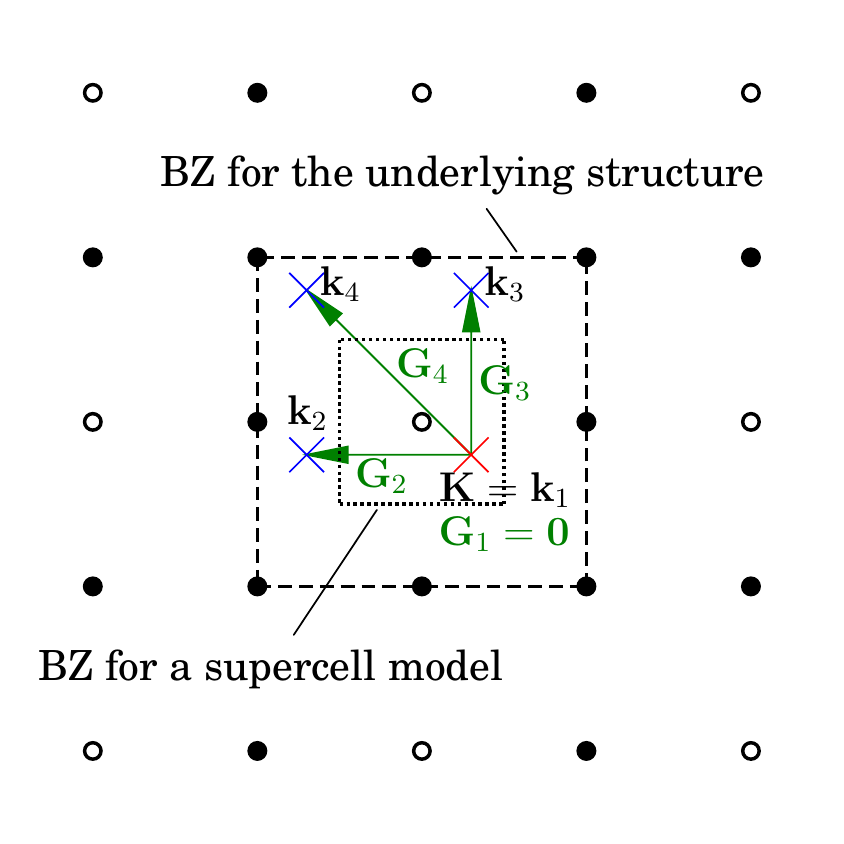}
\caption{
    (Color online)
    Two-dimensional representation of the relation 
    between a supercell model and its underlying structure
    in reciprocal space.
    White and black circles represent reciprocal lattice points of the
    supercell model,
    and the white circles also correspond to reciprocal lattice points of the
    underlying structure.
    A wave vector $\mathbf{K}$ moves to
    the wave vectors $\mathbf{k}_{i}$ inside the BZ for the underlying structure
    by adding the corresponding reciprocal lattice vectors 
    of the supercell model $\mathbf{G}_{i}$, represented by green arrows.
    %
    \label{fig:reciprocal_space}
}
\end{figure}

The basis of the lattice of a supercell model $\mathbf{A}_{i}$ can be constructed 
from the basis of the lattice of the underlying structure $\mathbf{a}_{i}$ as
$\mathbf{A}_{i} = \sum_{j = 1}^{3} n_{ji} \mathbf{a}_{j} \, (i = 1, 2, 3)$,
where 
$(n_{ji})$ 
is a $3 \times 3$ integer matrix.
The ``size'' of the supercell model relative to the underlying crystal structure
is then given as $N \equiv |\det (n_{ji})|$.
Lattice vectors of the underlying structure $\mathbf{t}$ and of the supercell
model $\mathbf{T}$ can be obtained as the integral linear combinations of 
$\mathbf{a}_{i}$ and $\mathbf{A}_{i}$, respectively.

The space group of the underlying structure is denoted as $\mathcal{G}$.
The point group of the space group $\mathcal{G}$,
which consists of the distinct rotational parts of the elements in $\mathcal{G}$,
is denoted as $\bar{\mathcal{G}}$.
The supercell model generally has lower crystallographic symmetry than $\mathcal{G}$.
The set of $\mathbf{t}$ forms the translation subgroup $\mathcal{T}$
of $\mathcal{G}$,
while the set of $\mathbf{T}$ forms the translation group
$\mathcal{T}'$,
which is a normal subgroup of $\mathcal{T}$.
$\mathcal{T}$ can be decomposed using the coset representatives relative to
$\mathcal{T}'$ as
\begin{align}
    \mathcal{T}
    &=
    \{ \mathbf{I}_{3} | \mathbf{t}_{1} \}
    \mathcal{T}'
    +
    \cdots
    +
    \{ \mathbf{I}_{3} | \mathbf{t}_{N} \}
    \mathcal{T}'
    ,
    \label{eq:coset_representatives_T}
\end{align}
where $\mathbf{I}_{3}$ is the $3 \times 3$ identity matrix.

The bases of the reciprocal lattice of the underlying structure $\mathbf{b}_{i}$ 
and of the supercell model $\mathbf{B}_{i}$ satisfy
$
\mathbf{a}_{i} \cdot \mathbf{b}_{j} 
= 
2 \pi \delta_{ij}
$
and
$
\mathbf{A}_{i} \cdot \mathbf{B}_{j} 
= 
2 \pi \delta_{ij},
$
respectively.
Reciprocal lattice vectors of the underlying structure $\mathbf{g}$ and 
of the supercell model $\mathbf{G}$ can be obtained as 
the integral linear combinations of $\mathbf{b}_{i}$ and $\mathbf{B}_{i}$, respectively.
Let $\{ \mathbf{g} \}$ be the set of $\mathbf{g}$.
$\mathbf{G}$ and satisfies the following relation;
\begin{align}
    \frac{1}{N}
    \sum_{j=1}^{N}
    e^{i \mathbf{G} \cdot \mathbf{t}_{j}}
    &=
    \begin{cases}
    1 & \textrm{if $\mathbf{G} \in \{ \mathbf{g} \}$,} \\
    0 & \textrm{otherwise,}
    \end{cases}
    \label{eq:fourier}
\end{align}
where the set of $\mathbf{t}_{j}$ is the lattice vectors of the underlying crystal structure 
corresponding to the coset representatives in Eq.~\eqref{eq:coset_representatives_T}.
Figure~\ref{fig:reciprocal_space} represents the relation of the first Brillouin zones (BZs) 
for the supercell model and for its underlying structure.
%
%
A wave vector $\mathbf{K}$ is related to $N$ distinct wave vectors 
$\mathbf{k}_{1}, \dots, \mathbf{k}_{N}$
inside the BZ for the underlying structure as
\begin{align}
    \mathbf{k}_{i} = \mathbf{K} + \mathbf{G}_{i}
    \quad (i = 1, \dots, N)
    ,
    \label{eq:wave_vectors_sc2pc}
\end{align}
where $\mathbf{G}_{i}$ is the reciprocal lattice vector of the supercell model 
corresponding to $\mathbf{k}_{i}$.
If $k = l$ then
$\mathbf{k}_{k} - \mathbf{k}_{l} = \mathbf{0} \in \{\mathbf{g}\}$,
while if $k \neq l$ then
$\mathbf{k}_{k} - \mathbf{k}_{l} \notin \{\mathbf{g}\}$
because both $\mathbf{k}_{k}$ and $\mathbf{k}_{l}$ are inside the BZ 
for the underlying structure.
Therefore, using Eq.~\eqref{eq:fourier},
\begin{align}
    \frac{1}{N}
    \sum_{j = 1}^{N}
    e^{i (\mathbf{k}_{k} - \mathbf{k}_{l}) \cdot \mathbf{t}_{j}}
    &=
    \delta_{kl}
    .
    \label{eq:fourier_k}
\end{align}

\subsection{Projection operators}

Three types of projection operators are derived 
to formulate the current band-unfolding method.
The projection operators for wave vectors $\hat{P}^{\mathbf{k}}$ are used 
to decompose the eigenvectors obtained from a supercell model 
according to the translational symmetry for the underlying crystal structure.
These projection operators are equivalent to those derived in 
Ref.~\cite{Allen2013, *Allen2013E}.
The projection operators for SRs $\hat{P}^{\mathbf{k}\mu}$ are used 
to further decompose the eigenvectors according to the SRs 
of little groups.
The projection operators for chemical elements $\hat{P}^{\textrm{X}}$ are also defined
to analyze the contributions of different combinations of the chemical elements 
to the unfolded band structures.

\subsubsection{Projection operators for wave vectors}
\label{sec:pos_wvs}

\begin{figure*}[tbp]
\centering
\includegraphics[width=0.85\linewidth]{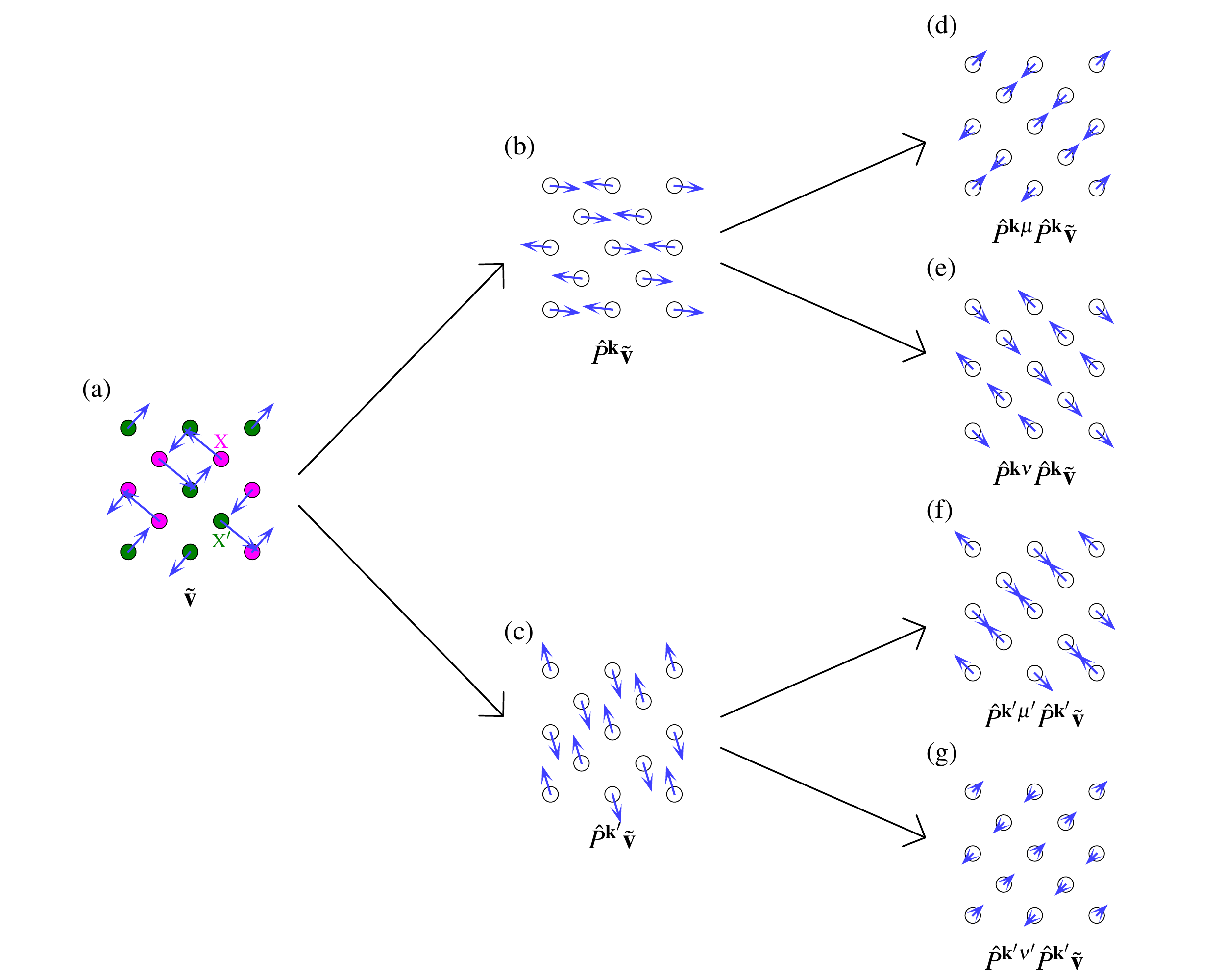}
\caption{
    (Color online)
    Two-dimensional representation of 
    how the projection operators for wave vectors 
    $\hat{P}^{\mathbf{k}}$ in Eq.~\eqref{eq:po_wvs}
    and for SRs
    $\hat{P}^{\mathbf{k}\mu}$ in Eq.~\eqref{eq:po_srs}
    work on a phonon mode eigenvector of a supercell model of a disordered system.
    Circles represent atoms in the system;
    red and green ones represent the chemical elements X and X$'$, respectively, 
    while white ones indicate that 
    the chemical elements are no longer distinguished.
    Blue arrows on the circles represent the real parts of a mode eigenvector
    on atoms or its projections.
    (a):
    Hypothetical phonon mode eigenvector $\tilde{\mathbf{v}}$.
    (b), (c):
    Projection of $\tilde{\mathbf{v}}$ by
    $\hat{P}^{\mathbf{k} }$ and
    $\hat{P}^{\mathbf{k}'}$ for different wave vectors $\mathbf{k}$ and $\mathbf{k}'$, 
    respectively.
    (d), (e), (f), (g):
    Further projection of
     $\hat{P}^{\mathbf{k} } \tilde{\mathbf{v}}$ 
    ($\hat{P}^{\mathbf{k}'} \tilde{\mathbf{v}}$)
    by
    $\hat{P}^{\mathbf{k}\mu}$ and $\hat{P}^{\mathbf{k}\nu }$ 
    ($\hat{P}^{\mathbf{k}'\mu'}$ and $\hat{P}^{\mathbf{k}'\nu'}$),
    where $\mu$ and $\nu$ ($\mu'$ and $\nu'$) are indices for the SRs 
    of $\mathcal{G}^{\mathbf{k}}$ ($\mathcal{G}^{\mathbf{k}'}$).
    \label{fig:po_srs}
}
\end{figure*}

Let $f^{\mathbf{K}} (\mathbf{x})$ be an eigenfunction 
obtained from the supercell model,
which transforms under the translations
$\hat{T} = \{ \mathbf{I}_{3} | \mathbf{T} \} \in \mathcal{T}'$ as
\begin{align}
\hat{T}
f^{\mathbf{K}}
(\mathbf{x})
&=
e^{-i \mathbf{K} \cdot \mathbf{T}}
f^{\mathbf{K}}
(\mathbf{x})
.
\label{eq:trans_K}
\end{align}
Note that both scalar-field and vector-field functions can be considered as 
$f^{\mathbf{K}} (\mathbf{x})$.
$f^{\mathbf{K}} (\mathbf{x})$ can be decomposed using the basis functions 
of the irreducible representations (IRs) of $\mathcal{T}$ as
\begin{align}
f^{\mathbf{K}} (\mathbf{x})
&=
\sum_{k = 1}^{N}
c_{\mathbf{k}_{k}}
f^{\mathbf{k}_{k}} (\mathbf{x})
,
\label{eq:decomposition_wvs}
\end{align}
where 
$\mathbf{k}_{k}$ ($k = 1, \dots, N$) is a wave vector 
inside the BZ for the underlying structure 
obtained from $\mathbf{K}$ according to Eq.~\eqref{eq:wave_vectors_sc2pc},
and $f^{\mathbf{k}_{k}} (\mathbf{x})$ is the basis function 
of the IR of $\mathcal{T}$ labeled $\mathbf{k}_{k}$.
By definition 
$f^{\mathbf{k}_{k}} (\mathbf{x})$
is transformed by the translations
$\hat{t} \equiv \{ \mathbf{I}_{3} | \mathbf{t} \} \in \mathcal{T}$ 
as
\begin{align}
    \hat{t}
    f^{\mathbf{k}_{k}} (\mathbf{x})
    &=
    f^{\mathbf{k}_{k}} (\hat{t}^{-1} \mathbf{x})
    \notag
    \\
    &=
    e^{-i\mathbf{k}_{k} \cdot \mathbf{t}}
    f^{\mathbf{k}_{k}} (\mathbf{x})
    ,
    \label{eq:transformation_wv}
\end{align}
where 
the set of $e^{- i \mathbf{k}_{k} \cdot \mathbf{t}}$
for all $\hat{t}$ is the IR labeled $\mathbf{k}_{k}$.
Note that all the IRs of $\mathcal{T}$ is one-dimensional 
because $\mathcal{T}$ is an Abelian group
\cite{Kim1999, Dresselhaus2007, ElBatanouny2008}.
$f^{\mathbf{k}_{k} } (\mathbf{x})$ and
$f^{\mathbf{k}_{k'}} (\mathbf{x})$ are orthogonal to each other
when $k \neq k'$.

The projection operator $\hat{P}^{\mathbf{k}_{k}}$
for the wave vector $\mathbf{k}_{k}$
can be constructed using the coset representatives 
$\hat{t}_{j} = \{ \mathbf{I}_{3} | \mathbf{t}_{j} \}$ of 
$\mathcal{T}$ relative to $\mathcal{T}'$ 
in Eq.~\eqref{eq:coset_representatives_T} as
\begin{align}
    \hat{P}^{\mathbf{k}_{k}}
    &=
    \frac{1}{N}
    \sum_{j = 1}^{N}
    \chi^{\mathbf{k}_{k}}
    (\hat{t}_{j})^{*}
    \hat{t}_{j}
    \notag
    \\
    &=
    \frac{1}{N}
    \sum_{j = 1}^{N}
    e^{i \mathbf{k}_{k} \cdot \mathbf{t}_{j}}
    \hat{t}_{j}
    ,
    \label{eq:po_wvs}
\end{align}
where 
$
\chi^{\mathbf{k}_{k}}
(\hat{t}_{j})
=
e^{-i\mathbf{k}_{k} \cdot \mathbf{t}_{j}}
$
is the character of 
$\hat{t}_{j}$
in the IR labeled $\mathbf{k}_{k}$.
Using the orthogonality relations in Eq.~\eqref{eq:fourier_k}
and the transformation rule for the basis functions of the IRs of $\mathcal{T}$ 
in Eq.~\eqref{eq:transformation_wv},
it can be shown that
\begin{align}
    \hat{P}^{\mathbf{k}_{k}}
    f^{\mathbf{K}} (\mathbf{x})
    &=
    c_{\mathbf{k}_{k}} f^{\mathbf{k}_{k}} (\mathbf{x})
    \label{eq:effect_of_pos_wvs_simple}
\end{align}
(see Appendix~\ref{sec:effect_of_pos_wvs} for detailed derivation).
Equation~\eqref{eq:effect_of_pos_wvs_simple} indicates that
$\hat{P}^{\mathbf{k}_{k}}$ extracts
from $f^{\mathbf{K}} (\mathbf{x})$
the basis function of the IR of $\mathcal{T}$ specified by the wave vector $\mathbf{k}_{k}$.
From Eqs.~\eqref{eq:decomposition_wvs} and~\eqref{eq:effect_of_pos_wvs_simple},
\begin{align}
    f^{\mathbf{K}} (\mathbf{x})
    &=
    \sum_{k = 1}^{N}
    \hat{P}^{\mathbf{k}_{k}}
    f^{\mathbf{K}} (\mathbf{x})
    .
    \label{eq:decomposition_wvs_op}
\end{align}
$\hat{P}^{\mathbf{k}_{k} } f^{\mathbf{K}} (\mathbf{x})$ and 
$\hat{P}^{\mathbf{k}_{k}'} f^{\mathbf{K}} (\mathbf{x})$ are orthogonal to each other
when $k \neq k'$.
Figures~\ref{fig:po_srs}(a)--(c) visualize how $\hat{P}^{\mathbf{k}_{k}}$
works on a phonon mode eigenvector of a supercell model.

\subsubsection{Projection operators for SRs}
\label{sec:pos_srs}

$f^{\mathbf{k}_{k}} (\mathbf{x})$
shows the translational symmetry of the underlying crystal structure
but in general is not a basis function of the SRs 
of the little group of the wave vector
$\mathbf{k}_{k}$.
Here the projection operators for SRs to decompose $f^{\mathbf{k}_{k}} (\mathbf{x})$ 
according to the SRs are derived.
For the sake of simplicity,
we hereafter omit 
the index for the wave vector $\mathbf{k}_{k}$ is hereafter omitted,
$f^{\mathbf{k}_{k}} (\mathbf{x})$ is denoted as $f^{\mathbf{k}} (\mathbf{x})$.

The little group $\mathcal{G}^{\mathbf{k}}$ is the subgroup of $\mathcal{G}$ 
whose elements $\{ \mathbf{R} | \mathbf{w} \}$ leave $\mathbf{k}$ invariant 
in the sense that $\mathbf{R}^{T} \mathbf{k} = \mathbf{k} + \mathbf{g}$
\cite{ITB_2010}.
The SRs $\Gamma^{\mathbf{k}\mu}$ of $\mathcal{G}^{\mathbf{k}}$ are defined as
the IRs of $\mathcal{G}^{\mathbf{k}}$ that satisfy
\begin{align}
    \Gamma^{\mathbf{k}\mu}
    (\{ \mathbf{I}_{3} | \mathbf{t} \})
    &=
    e^{-i \mathbf{k} \cdot \mathbf{t}}
    \mathbf{I}_{d_{\mu}}
    ,
    \label{eq:def_srs}
\end{align}
where $\mu$ is the index for the SRs,
$d_{\mu}$ is the dimension of the $\mu$th SR,
and $\mathbf{I}_{d_{\mu}}$ is the $d_{\mu} \times d_{\mu}$ identity matrix
\cite{Kim1999, Dresselhaus2007, ElBatanouny2008}.
Note that the number of the inequivalent SRs of $\mathcal{G}^{\mathbf{k}}$ is finite.

Since the SRs of $\mathcal{G}^{\mathbf{k}}$ satisfy Eq.~\eqref{eq:def_srs},
$f^{\mathbf{k}} (\mathbf{x})$ can be decomposed as
\begin{align}
    f^{\mathbf{k}} (\mathbf{x})
    &=
    \sum_{\mu}
    \sum_{s = 1}^{n_{\mu}}
    \sum_{k = 1}^{d_{\mu}}
    c_{\mu s k}
    f^{\mathbf{k} \mu s k} (\mathbf{x})
    ,
    \label{eq:decomposition_SRs}
\end{align}
where 
$
f^{\mathbf{k} \mu s k}
(\mathbf{x})
$
is the $k$th-row basis function of the $\mu$th SR belonging to the $s$th set for the SR,
$n_{\mu}$ is the number of the sets for the $\mu$th SR,
and
$c_{\mu s k}$ is the coefficient of the linear combination.
%
%
Note that there can be two or more sets of basis functions for the same SR.
By definition $f^{\mathbf{k} \mu s k} (\mathbf{x})$ is transformed by
$\hat{g} = \{ \mathbf{R} | \mathbf{w} \} \in \mathcal{G}^{\mathbf{k}}$ as
\begin{align}
    \hat{g}
    f^{\mathbf{k} \mu s k}
    (\mathbf{x})
    &=
    f^{\mathbf{k} \mu s k}
    (\hat{g}^{-1} \mathbf{x})
    \notag
    \\
    &=
    \sum_{k' = 1}^{d_{\mu}}
    f^{\mathbf{k} \mu s k'}
    (\mathbf{x})
    \Gamma^{\mathbf{k} \mu}_{k' k}
    (\hat{g})
    .
    \label{eq:transformation_sr}
\end{align}
$f^{\mathbf{k}  \mu s  k } (\mathbf{x})$ and
$f^{\mathbf{k}' \nu s' k'} (\mathbf{x})$ are orthogonal to each other
when $\mu \neq \nu$, $s \neq s'$, or $k \neq k'$.

$\mathcal{G}^{\mathbf{k}}$ can be decomposed using the coset representatives 
relative to $\mathcal{T}$ as 
\begin{align}
    \mathcal{G}^{\mathbf{k}}
    &=
    \{ \mathbf{R}_{1} | \mathbf{w}_{1} \}
    \mathcal{T}
    +
    \cdots
    +
    \{ \mathbf{R}_{|\lcg|} | \mathbf{w}_{|\lcg|} \}
    \mathcal{T},
    \label{eq:coset_representatives_Gk}
\end{align}
where and $\lcg$ is the little cogroup,
i.e.,
the point group composed of the distinct rotational parts of the elements
in $\mathcal{G}^{\mathbf{k}}$.
The projection operator $\hat{P}^{\mathbf{k}\mu}$
for the $\mu$th SR is constructed using 
$\hat{g}_{j} = \{ \mathbf{R}_{j} | \mathbf{w}_{j} \}$
as
\begin{align}
    \hat{P}^{\mathbf{k} \mu}
    &=
    \frac{d_{\mu}}{| \lcg |}
    \sum_{j = 1}^{| \lcg |}
        \chi^{\mathbf{k}\mu}
    (\hat{g}_{j})^{*}
    \hat{g}_{j}
    ,
    \label{eq:po_srs}
\end{align}
where 
$
    \chi^{\mathbf{k}\mu} (\hat{g}_{j})
    \equiv
    \mathrm{tr}\,[\Gamma^{\mathbf{k}\mu} (\hat{g}_{j})]
    =
    \sum_{m = 1}^{d_{\mu}}
    \Gamma^{\mathbf{k}\mu}_{mm}
    (\hat{g}_{j})
$
is the character of $\hat{g}_{j}$ in the $\mu$th SR.
Using the orthogonality relations for SRs in Eq.~\eqref{eq:orthogonality_sr}
(see Appendix~\ref{sec:orthogonality_relations_for_SRs})
and the transformation rule for the basis functions of the SRs
in Eq.~\eqref{eq:transformation_sr},
it can be shown that
\begin{align}
    \hat{P}^{\mathbf{k}\mu}
    f^{\mathbf{k}} (\mathbf{x})
    &=
    \sum_{s = 1}^{n_{\mu}}
    \sum_{k = 1}^{d_{\mu}}
    c_{\mu s k}
    f^{\mathbf{k} \mu s k}
    (\mathbf{x})
    \label{eq:effect_of_pos_SRs_simple}
\end{align}
(see Appendix~\ref{sec:effect_of_projection_operators_sr}
for detailed derivation).
Equation~\eqref{eq:effect_of_pos_SRs_simple} indicates that
$\hat{P}^{\mathbf{k}\mu}$ extracts
from $f^{\mathbf{k}} (\mathbf{x})$
the part being in the partial space spanned by the basis functions for the $\mu$th SR 
of $\mathcal{G}^{\mathbf{k}}$.
From Eqs.~\eqref{eq:decomposition_SRs} and~\eqref{eq:effect_of_pos_SRs_simple},
\begin{align}
    f^{\mathbf{k}} (\mathbf{x})
    &=
    \sum_{\mu}
    \hat{P}^{\mathbf{k}\mu}
    f^{\mathbf{k}} (\mathbf{x})
    .
    \label{eq:decomposition_SRs_op}
\end{align}
$\hat{P}^{\mathbf{k}\mu} f^{\mathbf{k}} (\mathbf{x})$ and 
$\hat{P}^{\mathbf{k}\nu} f^{\mathbf{k}} (\mathbf{x})$
are orthogonal to each other when $\mu \neq \nu$.
Figures~\ref{fig:po_srs}(d)--(g) visualize how $\hat{P}^{\mathbf{k}\mu}$
works on a phonon mode eigenvector of a supercell model.

Practically $\hat{P}^{\mathbf{k}\mu}$ can be explicitly obtained as follows.
As shown in Appendix~\ref{sec:orthogonality_relations_for_SRs},
the SRs of $\mathcal{G}^{\mathbf{k}}$ can be constructed 
using the irreducible projective representations (IPRs) of $\lcg$.
Since the characters of the IPRs of $\lcg$ are tabulated in the literature
\cite{ITD_2013},
the characters of the SRs of $\mathcal{G}^{\mathbf{k}}$ can be obtained from these data
using Eq.~\eqref{eq:small_rep_and_proj_rep}.
$\hat{P}^{\mathbf{k}\mu}$ is then calculated from the obtained characters 
of the SRs of $\mathcal{G}^{\mathbf{k}}$.

It should be emphasized that 
from a mathematical viewpoint the derivations of $\hat{P}^{\mathbf{k}}$ and 
of $\hat{P}^{\mathbf{k}\mu}$ are very similar.
The only difference is that for $\hat{P}^{\mathbf{k}\mu}$ 
two- or more-dimensional SRs have to be dealt with.

\subsubsection{Projection operators for chemical elements}
\label{sec:pos_chemical_elements}

A phonon mode eigenvector
$\tilde{\mathbf{v}}$
(here $\tilde{\mathbf{v}} (\mathbf{K}, J)$ is simply denoted as $\tilde{\mathbf{v}}$)
of a supercell model can be decomposed into the contributions from different elements  
using the projection operator $\hat{P}^{\textrm{X}}$ for the chemical element X,
which works on $\tilde{mathbf{v}}$ as
\begin{align}
    [
    \hat{P}^{\textrm{X}}
    \tilde{\mathbf{v}}
    ]_{l \kappa}
    &=
    \begin{cases}
    \tilde{\mathbf{v}}_{l \kappa}
    &
    \textrm{if X is on the site $l \kappa$},
    \\
    0
    &
    \textrm{otherwise}.
    \end{cases}
    \label{eq:pos_chemical_elements}
\end{align}
$\hat{P}^{\textrm{X}}$ satisfies
\begin{align}
    \tilde{\mathbf{v}}
    &=
    \sum_{\textrm{X}}
    \hat{P}^{\textrm{X}}
    \tilde{\mathbf{v}}
    ,
\end{align}
where the summation is taken over all the chemical elements in the system.
When $\textrm{X} \neq \textrm{X}'$,
$\hat{P}^{\textrm{X} }\tilde{\mathbf{v}}$
and
$\hat{P}^{\textrm{X}'}\tilde{\mathbf{v}}$
are orthogonal to each other.
However,
$\hat{P}^{\mathbf{k}}\hat{P}^{\textrm{X} }\tilde{\mathbf{v}}$
and
$\hat{P}^{\mathbf{k}}\hat{P}^{\textrm{X}'}\tilde{\mathbf{v}}$
are not necessarily orthogonal to each other,
as well as
$\hat{P}^{\mathbf{k}\mu} \hat{P}^{\mathbf{k}}\hat{P}^{\textrm{X}}
\tilde{\mathbf{v}}$
and
$\hat{P}^{\mathbf{k}\mu} \hat{P}^{\mathbf{k}}\hat{P}^{\textrm{X}'}
\tilde{\mathbf{v}}$
are not necessarily orthogonal to each other.

\begin{figure*}[tbp]
\centering
\includegraphics[width=0.85\linewidth]{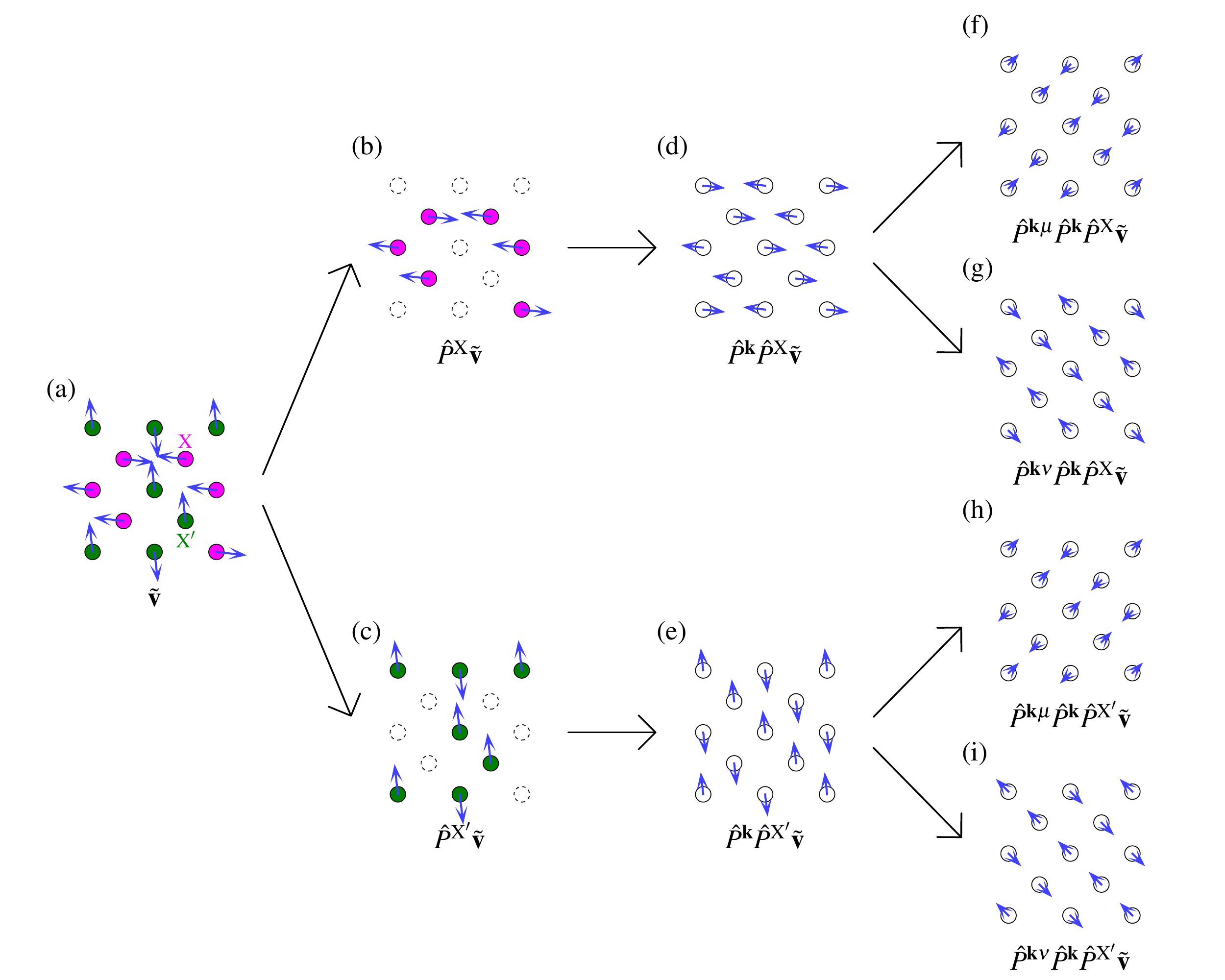}
\caption{
    (Color online)
    Two-dimensional representation of how the projection operators for chemical elements 
    $\hat{P}^{\textrm{X}}$
    in Eq.~\eqref{eq:pos_chemical_elements}
    work on a phonon mode eigenvector of a supercell model of a disordered
    system.
    The symbols are used in the same way as those in Fig.~\ref{fig:po_srs}.
    (a):
    Hypothetical mode eigenvector $\tilde{\mathbf{v}}$.
    (b), (c):
    Projection of $\tilde{\mathbf{v}}$ by
    $\hat{P}^{\textrm{X}}$
    and
    $\hat{P}^{\textrm{X}'}$.
    (d), (e):
    Further projection of
    $\hat{P}^{\textrm{X} } \tilde{\mathbf{v}}$ 
    and
    $\hat{P}^{\textrm{X}'} \tilde{\mathbf{v}}$
    by
    $\hat{P}^{\mathbf{k}}$.
    (f), (g), (h), (i):
    Further projection of 
    $\hat{P}^{\mathbf{k} } \hat{P}^{\textrm{X} } \tilde{\mathbf{v}}$ 
    and
    $\hat{P}^{\mathbf{k} } \hat{P}^{\textrm{X}'} \tilde{\mathbf{v}}$
    by 
    $\hat{P}^{\mathbf{k}\mu}$ and $\hat{P}^{\mathbf{k}\nu}$.
    \label{fig:po_chemical_elements}
}
\end{figure*}

Figure~\ref{fig:po_chemical_elements} visualizes how
$\hat{P}^{\textrm{X}}$
and
$\hat{P}^{\textrm{X}'}$
for two different chemical components X and X$'$, respectively,
works on $\tilde{\mathbf{v}}$.
In this figure,
$\hat{P}^{\mathbf{k}\mu} \hat{P}^{\mathbf{k}} \hat{P}^{\textrm{X} }
\tilde{\mathbf{v}}$ 
and 
$\hat{P}^{\mathbf{k}\mu} \hat{P}^{\mathbf{k}} \hat{P}^{\textrm{X}'}
\tilde{\mathbf{v}}$ 
point the same direction for each atomic position.
Those projected vectors are 
``positively correlated''
in the sense that the real part of the dot product between
$
\hat{P}^{\mathbf{k}\mu} \hat{P}^{\mathbf{k}} \hat{P}^{\textrm{X} }
\tilde{\mathbf{v}}
$
and
$
\hat{P}^{\mathbf{k}\mu} \hat{P}^{\mathbf{k}} \hat{P}^{\textrm{X}'}
\tilde{\mathbf{v}}
$
is positive.
In contrast,
$\hat{P}^{\mathbf{k}\nu} \hat{P}^{\mathbf{k}} \hat{P}^{\textrm{X} }
\tilde{\mathbf{v}}$ 
and 
$\hat{P}^{\mathbf{k}\nu} \hat{P}^{\mathbf{k}} \hat{P}^{\textrm{X}'}
\tilde{\mathbf{v}}$ 
point the opposite direction for each atomic position.
Those projected vectors are 
``negatively correlated''
in the sense that the real part of the dot product between
$
\hat{P}^{\mathbf{k}\nu} \hat{P}^{\mathbf{k}} \hat{P}^{\textrm{X} }
\tilde{\mathbf{v}}
$
and
$
\hat{P}^{\mathbf{k}\nu} \hat{P}^{\mathbf{k}} \hat{P}^{\textrm{X}'}
\tilde{\mathbf{v}}
$
is negative.
In this case,
each of X and X$'$ contributes to the $\nu$th SR in itself for
$\tilde{\mathbf{v}}$,
but in total they cancel out each other.


\subsection{Spectral functions}
\label{sec:spectral_functions}

Here we obtain the spectral functions,
which are regarded as the unfolded band structures.
We use the notations for phonon modes in
Sec.~\ref{sec:phonon_mode_calculations}
for the sake of simplicity.

\subsubsection{Spectral functions at each $\mathbf{k}$}

Let us first consider the ``original'' spectral function
of a supercell model
$A_s (\mathbf{K}, \omega)$
as
\begin{align}
    A_s (\mathbf{K}, \omega)
    &=
    \sum_{J}
    \delta [\omega - \omega (\mathbf{K}, J)]
    .
    \label{eq:original_sf_1}
\end{align}
The peak positions of the delta functions constitute
the ``original'' band structure of the supercell model.

The \textit{unfolded} spectral function
$A (\mathbf{k}_{k}, \omega)$ is defined 
using the projection operators for wave vectors 
$\hat{P}^{\mathbf{k}_{k}}$ in Eq.~\eqref{eq:po_wvs} as
\begin{align}
    A (\mathbf{k}_{k}, \omega)
    &\equiv
    \sum_{J}
    \left|
    [\hat{P}^{\mathbf{k}_{k} } \tilde{\mathbf{v}} (\mathbf{K}, J)]_{l}
    \right|^{2}
    \delta [\omega - \omega (\mathbf{K}, J)]
    ,
    \label{eq:total_sf}
\end{align}
where $A (\mathbf{k}_{k}, \omega)$ does not depends on 
the choice of the index for supercells $l$.
$A (\mathbf{k}_{k}, \omega)$ satisfies
\begin{align}
A_s (\mathbf{K}, \omega)
&=
\sum_{k=1}^{N}
A (\mathbf{k}_{k}, \omega)
,
\end{align}
where
we use Eq.~\eqref{eq:decomposition_wvs_op} and the orthogonality between
$[\hat{P}^{\mathbf{k}_{k }} \tilde{\mathbf{v}} (\mathbf{K}, J)]_{l}$
and
$[\hat{P}^{\mathbf{k}_{k'}} \tilde{\mathbf{v}} (\mathbf{K}, J)]_{l}$
when
$k \neq k'$.
It can be said that
$A_{s} (\mathbf{K}, \omega)$ defined in the BZ for the supercell model
is remapped in the BZ for the underlying crystal structure
with the weights obtained from
$[\hat{P}^{\mathbf{k}} \tilde{\mathbf{v}} (\mathbf{K}, J)]_{l}$.
$A (\mathbf{k}_{k}, \omega)$ is equivalent to the unfolded spectral function in previous reports
\cite{Popescu2012, Allen2013, *Allen2013E}.


The partial spectral function
$A^{\mu} (\mathbf{k}, \omega)$ (the index for $\mathbf{k}$ is hereafter omitted for the sake of simplicity) for the $\mu$th SR of $\mathcal{G}^{\mathbf{k}}$
is defined using the projection operator for SRs
$\hat{P}^{\mathbf{k}\mu}$ in Eq.~\eqref{eq:po_srs} as
\begin{align}
    A^{\mu} (\mathbf{k}, \omega)
    &\equiv
    \sum_{J}
    \left|
    [\hat{P}^{\mathbf{k}\mu} \hat{P}^{\mathbf{k}} \tilde{\mathbf{v}} (\mathbf{K}, J)]_{l}
    \right|^{2}
    \delta [\omega - \omega (\mathbf{K}, J)]
    .
    \label{eq:partial_sf_s}
\end{align}
$A^{\mu} (\mathbf{k}, \omega)$ satisfies
\begin{align}
    A (\mathbf{k}, \omega)
    &=
    \sum_{\mu}
    A^{\mu} (\mathbf{k}, \omega)
    ,
    \label{eq:total_sf_to_partial_sf_s}
\end{align}
where we use Eq.~\eqref{eq:decomposition_SRs_op}
and the orthogonality between
$[\hat{P}^{\mathbf{k}\mu} \hat{P}^{\mathbf{k}} \tilde{\mathbf{v}} (\mathbf{K}, J)]_{l}$
and
$[\hat{P}^{\mathbf{k}\nu} \hat{P}^{\mathbf{k}} \tilde{\mathbf{v}} (\mathbf{K}, J)]_{l}$
when $\mu \neq \nu$.

$A^{\mu} (\mathbf{k}, \omega)$ can be further decomposed
using the projection operators for chemical elements
$\hat{P}^{\textrm{X}}$
in Eq.~\eqref{eq:pos_chemical_elements} as
\begin{align}
    A^{\mu} (\mathbf{k}, \omega)
    &=
    \sum_{\textrm{X}, \textrm{X}'}
    A^{\mu, \textrm{XX}'} (\mathbf{k}, \omega)
    ,
\end{align}
where
\begin{multline}
    A^{\mu, \textrm{XX}'} (\mathbf{k}, \omega)
    \\
    \equiv
    \sum_{J}
    [\hat{P}^{\mathbf{k}\mu} \hat{P}^{\mathbf{k}} \hat{P}^{\textrm{X}}
    \tilde{\mathbf{v}} (\mathbf{K}, J)
    ]_{l}^{\dagger}
    [\hat{P}^{\mathbf{k}\mu} \hat{P}^{\mathbf{k}} \hat{P}^{\textrm{X}'}
    \tilde{\mathbf{v}} (\mathbf{K}, J)
    ]_{l} 
    \delta [\omega - \omega (\mathbf{K}, J)]
    .
    \\
    \label{eq:partial_sf_s_e}
\end{multline}
$A^{\mu, \textrm{XX}'} (\mathbf{k}, \omega)$
represents the contribution of the combination of the elements X and X$'$
to $A^{\mu} (\mathbf{k}, \omega)$.
$A^{\mu, \textrm{XX}} (\mathbf{k}, \omega)$
is for the contribution only from X,
while
$
A^{\mu, \overline{\textrm{XX$'$}}} (\mathbf{k}, \omega)
\equiv
A^{\mu, \textrm{XX$'$}} (\mathbf{k}, \omega) + A^{\mu, \textrm{X$'$X}}
(\mathbf{k}, \omega)\, (\textrm{X} = \textrm{X}')
$
is for the correlative contribution from X and X$'$.
$A^{\mathbf{k}\mu, \textrm{X}\textrm{X}} (\mathbf{k}, \omega)$
is always non-negative,
while
$A^{\mathbf{k}\mu, \overline{\textrm{XX$'$}}} (\mathbf{k}, \omega)$
($\textrm{X} \neq \textrm{X}'$)
becomes negative 
when the atomic movements of $\textrm{X}$ and $\textrm{X}'$ for the $\mu$th SR 
are negatively correlated as described in Sec.~\ref{sec:pos_chemical_elements}
and Fig.~\ref{fig:po_chemical_elements}.

\subsubsection{Average spectral functions over crystallographically equivalent wave vectors} 
\label{sec:star}

Although we can obtain the spectral functions using the procedure described above,
generally they still do not fully show the rotational symmetry 
for the underlying crystal structure.
To impose the crystallographic symmetry of the underlying crystal structure 
to the spectral functions obtained using the band-unfolding method,
we take the average of the spectral functions over the wave vectors 
that are crystallographically equivalent 
for the underlying structure
in the same manner as described in Sec.~III E in Ref.~\cite{Popescu2012}.

For a wave vector $\mathbf{k}$, the set of crystallographically equivalent wave vectors $\{ \mathbf{k} \}$ is called the star of $\mathbf{k}$
\cite{ITB_2010, ITD_2013}.
The average spectral function $\bar{A} (\mathbf{k}, \omega)$ at $\mathbf{k}$ 
is calculated as
\begin{align}
    \bar{A} 
    (\mathbf{k}, \omega)
    &=
    \frac{1}{| \{ \mathbf{k} \} |}
    \sum_{\mathbf{k}' \in \{ \mathbf{k} \}}
    A
    (\mathbf{k}', \omega).
    \label{eq:total_sf_average}
\end{align}
Similarly,
we can take the average also for 
$A^{\mu} (\mathbf{k}, \omega)$,
$A^{\mu, \textrm{XX}} (\mathbf{k}, \omega)$,
and
$A^{\mu, \overline{\textrm{XX$'$}}} (\mathbf{k}, \omega)$
as
\begin{gather}
\bar{A}^{\mu}
(\mathbf{k}, \omega)
=
    \frac{1}{| \{ \mathbf{k} \} |}
    \sum_{\mathbf{k}' \in \{ \mathbf{k} \}}
A^{\mu}
(\mathbf{k}', \omega),
\label{eq:partial_sf_s_average}
\\
\bar{A}^{\mu, \textrm{X}\textrm{X}}
(\mathbf{k}, \omega)
=
    \frac{1}{| \{ \mathbf{k} \} |}
    \sum_{\mathbf{k}' \in \{ \mathbf{k} \}}
A^{\mu, \textrm{X}\textrm{X}}
(\mathbf{k}', \omega),
\label{eq:partial_sf_s_e_1_average}
\end{gather}
and
\begin{gather}
\bar{A}^{\mu, \overline{\textrm{XX$'$}}}
(\mathbf{k}, \omega)
=
    \frac{1}{| \{ \mathbf{k} \} |}
    \sum_{\mathbf{k}' \in \{ \mathbf{k} \}}
A^{\mu, \overline{\textrm{XX$'$}}}
(\mathbf{k}', \omega),
\label{eq:partial_sf_s_e_2_average}
\end{gather}
respectively.

\section{COMPUTATIONAL DETAILS}
\label{sec:computational_details}

Here the computational details to obtain the effective phonon band structure 
of disordered fcc Cu$_{0.75}$Au$_{0.25}$
using the current band-unfolding method are described.

\subsection{Supercell models of disordered fcc Cu$_{0.75}$Au$_{0.25}$}
\label{sec:models_of_the_disordered_alloys}

The atomic configuration
in disordered fcc Cu$_{0.75}$Au$_{0.25}$ 
was approximated
using special quasirandom structures (SQSs)
\cite{Zunger1990}.
SQSs mimic fully disordered atomic configurations within limited-size supercells 
in terms of the correlation functions in the cluster expansion method
\cite{Sanchez1984,Fontaine1994,Ducastelle1994}.
In this study
the SQSs for the $2 \times 2 \times 2$ (32 atoms) and the
$3 \times 3 \times 3$ (108 atoms) supercells of the conventional fcc unit cell
were constructed and used to model disordered fcc Cu$_{0.75}$Au$_{0.25}$.
The SQSs were obtained using simulated annealing
\cite{Kirkpatrick1983, Kirkpatrick1984}
as implemented in the \textsc{CLUPAN} code
\cite{Seko2006, Seko2009}.

\subsection{Electronic structures}
\label{sec:electronic_structures}

The plane-wave basis projector augmented wave (PAW) method
\cite{Blochl1994} 
was employed in the framework of
density-functional theory
within the generalized gradient approximation
of the Perdew-Burke-Ernzerhof form
\cite{Perdew1996} 
as implemented in the \textsc{VASP} code
\cite{Kresse1995, Kresse1996, Kresse1999}.
A plane-wave energy cutoff of 350~eV was used.
$3d$ and $4s$ electrons were treated as valence electrons for Cu,
and
$5d$ and $6s$ electrons were treated as valence electrons for Au.
Other electrons were kept frozen.
The BZs were sampled by the $\Gamma$-centered $12\times12\times12$
$k$-point mesh per conventional fcc unit cell,
and the Methfessel-Paxton scheme~\cite{Methfessel1989}
with a smearing width of 0.4~eV was employed.
The total energies were minimized until the energy convergences were
less than 10$^{-8}$~eV.
Lattice shapes were kept to be cubic,
and lattice constants of Cu$_{0.75}$Au$_{0.25}$ were fixed to the experimental
value at room temperature,
3.753~{\AA}
\cite{Pearson1958}.
Atoms in the supercell models were initially put on the fcc atomic sites,
and then the internal atomic positions were optimized 
until the residual forces became less than $1 \times 10^{-3}$~eV/\AA.

\subsection{Band-unfolding for phonons}
\label{sec:band_unfolding_for_phonons}

The unfolded phonon band structure of disordered fcc Cu$_{0.75}$Au$_{0.25}$ 
was obtained as follows.
First the second-order force constants of the supercell models were calculated 
by applying finite atomic displacements of 0.01~{\AA} to the supercell models
with the optimized internal atomic positions.
No further expansion of the supercell models was applied to calculate the force
constants in this study.
Next the phonon modes of the supercell models were obtained 
according to Sec.~\ref{sec:phonon_mode_calculations}.
Here the atoms in the supercell models were supposed to be 
exactly on the fcc atomic sites before the optimization
of the internal atomic positions
to provide one-to-one correspondence for the atomic positions 
between the supercell models and their underlying fcc structure.
Finally the phonon modes were unfolded into the BZ for the primitive fcc unit cell;
total and partial spectral functions 
described in Eqs.~\eqref{eq:total_sf_average}, \eqref{eq:partial_sf_s_average}, 
\eqref{eq:partial_sf_s_e_1_average}, and \eqref{eq:partial_sf_s_e_2_average}
were calculated.
Delta functions in these spectral functions were smeared 
by the Lorentzian functions with the half-width at half-maximum of 0.05~THz for plotting.
The band-unfolding was performed using our own script
in combination with the \textsc{PHONOPY} code~\cite{Togo2015, Togo2008}.

\section{RESULTS AND DISCUSSION}
\label{sec:results}


\begin{figure}[tbp]
\centering
\includegraphics[width=\linewidth]{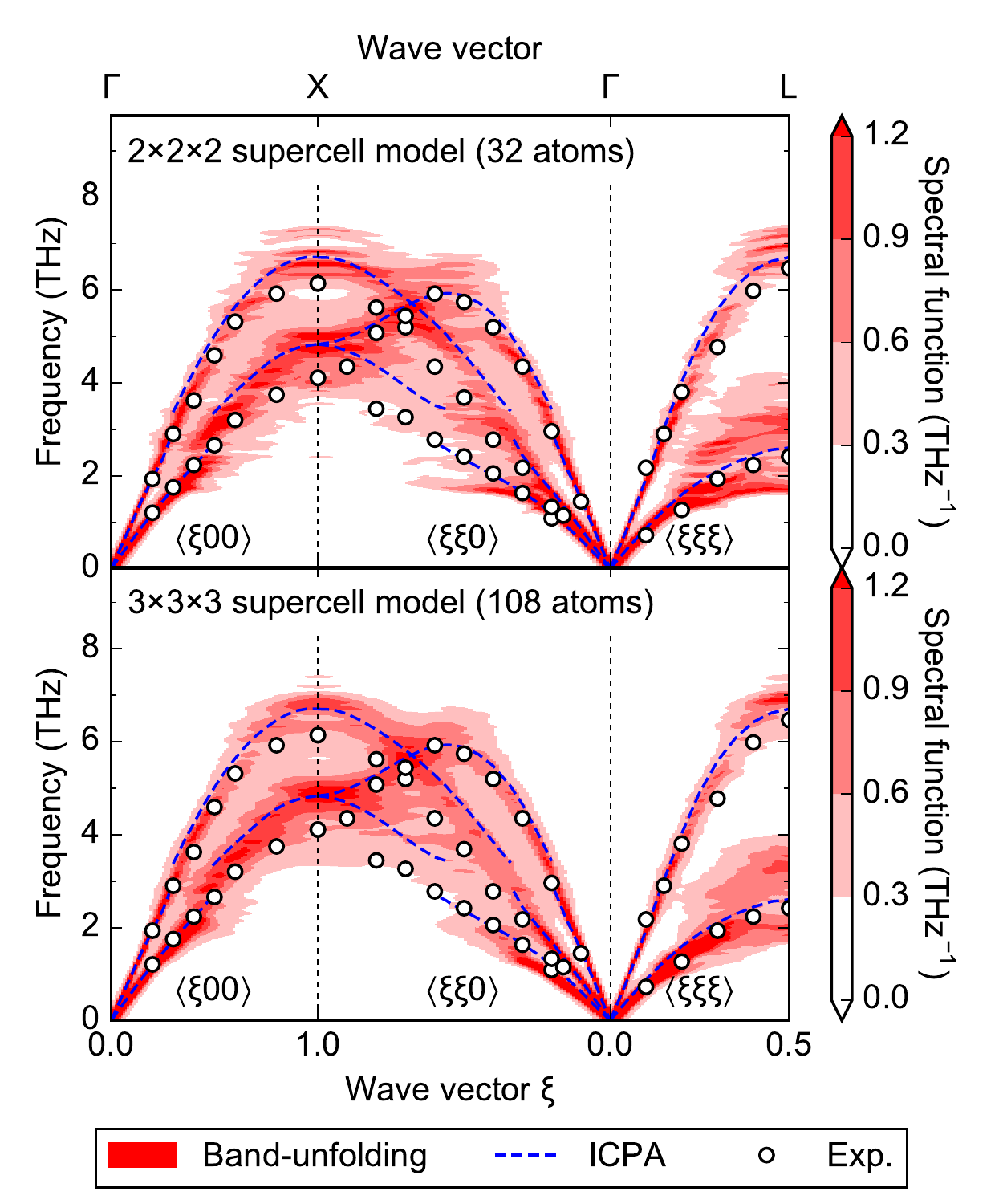}
\caption{
(Color online)
Phonon band structure of disordered fcc Cu$_{0.75}$Au$_{0.25}$
calculated using the band-unfolding method.
The upper and the lower panels show the results obtained from 
the $2 \times 2 \times 2$ and $3 \times 3 \times 3$ supercell models, respectively.
White circles represent experimental data at room temperature
\cite{Katano1988},
and blue dashed curves represent the result calculated using the ICPA method
\cite{Dutta2010}.
\label{fig:phonons_Cu075Au025}
}
\end{figure}

\begin{figure*}[tbp]
\centering
\includegraphics[height=13.5cm]{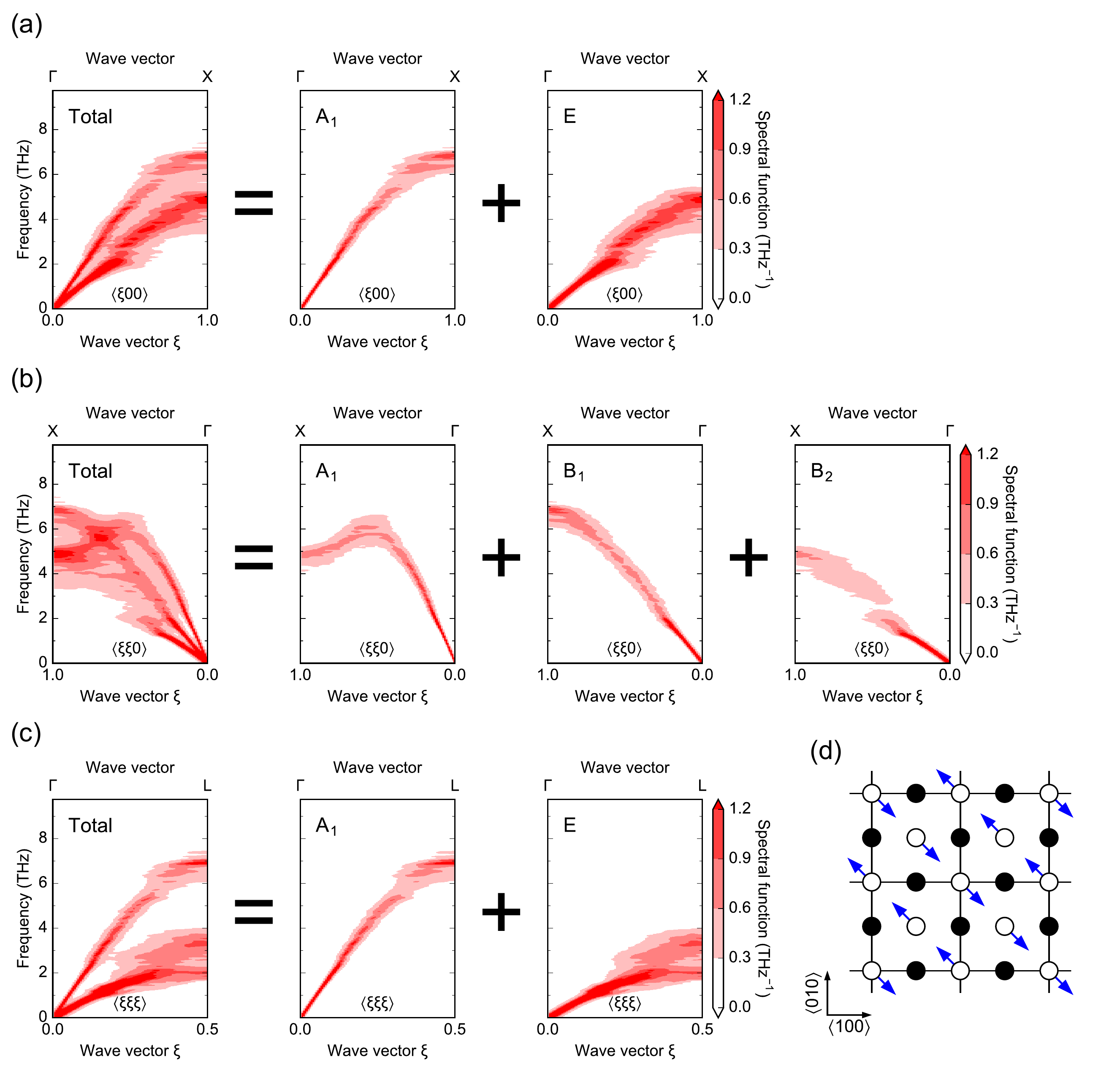}
\caption{
(Color online)
(a):
Decomposition of the unfolded phonon band structure of disordered fcc
Cu$_{0.75}$Au$_{0.25}$ according to the SRs
along the $\langle 100 \rangle$ direction.
The result is obtained from the $3 \times 3 \times 3$ supercell model.
The leftmost panel shows the total spectral function,
while the other panels show the partial spectral functions corresponding to
different SRs specified at the upper-left of the panels.
(b):
The same as (a) but along the $\langle 110 \rangle$ direction.
(c):
The same as (a) but along the $\langle 111 \rangle$ direction.
(d):
Atomic displacements of the $B_2$ mode at the wave vector
$\langle 0.5, 0.5, 0.0 \rangle$ at a certain moment.
Squares represent the conventional fcc unit cells,
white and black circles represent the atoms in different layers along the
$\langle 001 \rangle$ direction,
and blue arrows represent the atomic displacements.
\label{fig:phonons_separated}
}
\end{figure*}

\begin{figure*}[tbp]
\centering
\includegraphics[height=9cm]{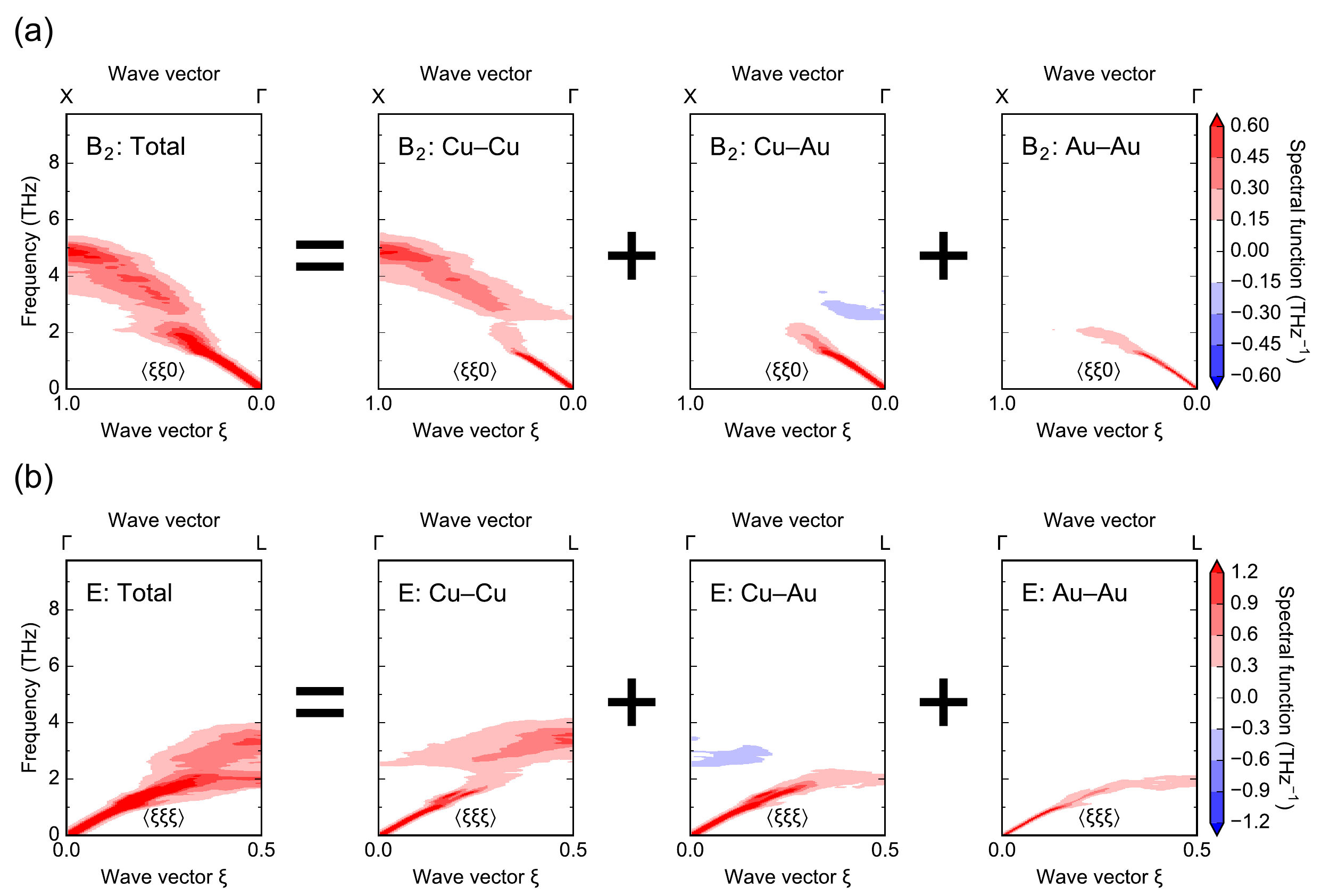}
\caption{
(Color online)
(a):
Decomposition of the partial spectral function for the $B_2$ mode 
along the $\langle 110 \rangle$ direction 
into the contributions of the combinations of the chemical elements 
for disordered fcc Cu$_{0.75}$Au$_{0.25}$.
The result is obtained from the $3 \times 3 \times 3$ supercell model.
The leftmost panel shows the partial spectral function for the $B_2$ mode in total,
while the other panels show the contributions of the combinations of chemical
elements specified at the upper-left in the panels.
(b):
The same as (a) but for the doubly-degenerated $E$ modes 
along the $\langle 111 \rangle$ direction.
\label{fig:phonons_chemical_contributions} 
}
\end{figure*}

%
%
%

Figure~\ref{fig:phonons_Cu075Au025} shows the phonon band structure
of disordered fcc Cu$_{0.75}$Au$_{0.25}$
obtained using the band-unfolding method
based on the first-principles calculations.
The $3 \times 3 \times 3$ supercell model shows much smoother spectral function
than the $2 \times 2 \times 2$ supercell model.
To investigate the convergence of the unfolded phonon band structure
with respect to supercell size in more detail,
the band-unfolding calculations are also performed 
using empirical interatomic potentials
(see Appendix~\ref{sec:empirical_potentials}).
It is found that the unfolded phonon band structure 
obtained from the $3 \times 3 \times 3$ supercell model is in excellent agreement with that 
obtained from the $6 \times 6 \times 6$ supercell model (864 atoms)
and that even the $2 \times 2 \times 2$ supercell model
gives the unfolded phonon band structure
qualitatively in good agreement with that 
obtained from the $6 \times 6 \times 6$ supercell model.
Hereafter the results obtained from the $3 \times 3 \times 3$ supercell model are focused on.


The peak positions of the spectral function roughly form the curves 
similar to the phonon band structures of typical pure fcc metals.
%
%
The unfolded phonon band structure, however, also shows the ``linewidths'' of phonon modes.
Since the current phonon modes are obtained under the harmonic approximation, 
these linewidths originate not from the phonon anharmonicity
but from the variations of atomic masses and force constants 
among the atomic sites due to the chemical disorder in Cu$_{0.75}$Au$_{0.25}$.
Actually, the atomic mass of Au relative to Cu is approximately 3.5.
Moreover, the values of the force constants in Cu$_{0.75}$Au$_{0.25}$ 
strongly depend on the combinations of the chemical elements and interatomic distance,
as shown in Appendix~\ref{sec:force_constants}.


The unfolded phonon band structure can be decomposed 
according to the SRs
as shown in Eq.~\eqref{eq:total_sf_to_partial_sf_s}
using the projection operators for SRs
defined in Eq.~\eqref{eq:po_srs}.
Figure~\ref{fig:phonons_separated} shows this decomposition
for the unfolded phonon band structure of
disordered fcc Cu$_{0.75}$Au$_{0.25}$.
%
The $A_{1}$ and the $B_{1}$ modes (in the Mulliken notation)
along the $\langle 110 \rangle$ direction are clearly separated
even when they cross to each other around the wave vector
$\langle 0.7, 0.7, 0.0 \rangle$.
This is difficult 
in a previous band-unfolding approach
\cite{Boykin2014},
where each mode in the unfolded band structure is determined 
based on the cumulative spectral function.
%
The modes obtained in the current band-unfolding method are associated with the SRs
and hence have the information on the crystallographic symmetry.
This enables us to analyze the unfolded band structures of disordered systems
in very similar manners to the ordinary band structures of ordered systems.


In Fig.~\ref{fig:phonons_separated},
the partial spectral functions clearly show several peculiar behaviors 
which cannot be found for pure metals or ordered alloys.
One is that
the $B_2$-mode branch
along the $\langle 110 \rangle$ direction,
whose atomic displacements are shown in Fig.~\ref{fig:phonons_separated}(d),
looks discontinuous around the wave vector 
$\langle 0.4, 0.4, 0.0 \rangle$.
At this wave vector,
there is a jump of the peak positions of the spectral function
from around 2~THz to around 3~THz.
Another peculiar behavior is that
the doubly-degenerated $E$-modes branch along the $\langle 111 \rangle$ direction 
looks split 
around the midpoint between the $\Gamma$ and the L points.
At the L point,
the split peak positions are found around 2~THz and around 3--4~THz.
To investigate whether these peculiar behaviors are found 
also in larger-size supercell models or not,
we also check the unfolded phonon band structure of Cu$_{0.75}$Au$_{0.25}$ 
calculated using empirical interatomic potentials
(see Appendix~\ref{sec:empirical_potentials}).
It is found that
the discontinuous and the split modes are still found
up to the $6 \times 6 \times 6$ supercell model (864~atoms), 
where the unfolded phonon band structure is almost converged with respect to the supercell size.
Therefore,
these peculiar behaviors are probably not spurious ones due to the limited supercell size 
but reveal physically meaningful characteristics
of disordered fcc Cu$_{0.75}$Au$_{0.25}$
originating from its chemical disorder.


To elucidate the origins of the discontinuous and the split branches,
we further decompose the partial spectral functions for these modes 
into the contributions of the combinations of the chemical elements 
as shown in Eqs.~\eqref{eq:partial_sf_s_e_1_average} and \eqref{eq:partial_sf_s_e_2_average}.
Figure~\ref{fig:phonons_chemical_contributions}(a) shows the result for the $B_2$ mode along the $\langle 110 \rangle$ diretion.
Around the $\Gamma$ point,
all Cu--Cu, Cu--Au, and Au--Au contribute to the $B_2$ modes.
Cu--Cu also contributes to the $B_2$ modes around 2--3~THz,
but this Cu--Cu contribution is canceled out by the negative Cu--Au contribution.
As explained in Sec.~\ref{sec:pos_chemical_elements} and Fig.~\ref{fig:po_chemical_elements},
when the Cu--Au contribution is negative,
Cu and Au atoms hypothetically on the same position
tend to move to the opposite directions,
although each chemical element in itself shows the $B_2$-mode atomic movements.
When the wave vector goes away from the $\Gamma$ point,
the peak contributed by all the combinations of the chemical elements almost disappear 
around the wave vector $\langle 0.4, 0.4, 0.0 \rangle$,
where the peak frequency is around 2~THz.
Instead, at this wave vector,
the Cu--Cu contribution makes a new peak around 3~THz,
which continues up to the X point.
As a result,
the $B_2$-mode branch looks discontinuous around 
$\langle 0.4, 0.4, 0.0 \rangle$.
Figure~\ref{fig:phonons_chemical_contributions}(b) shows the 
decomposition of the doubly-degenerated $E$ modes along the $\langle 111 \rangle$ direction into the contributions of the combinations of the chemical elements.
Like the $B_2$ mode along the $\langle 110 \rangle$ direction,
all Cu--Cu, Cu--Au, and Au--Au contribute to the $E$ modes around the $\Gamma$
point.
We can also find the cancellation between the Cu--Cu and the Cu--Au contributions around 2--3~THz.
Cu--Au and Au--Au keep to contribute to this peak up to the L points,
while Cu--Cu less contribute to this peak as the wave vectors goes away from the $\Gamma$ point.
Instead, around the midpoint between the $\Gamma$ and the L points,
Cu--Cu makes a new peak around 3~THz,
which continues up to the L point.
As a result,
we can find two peaks from around the midpoint between the $\Gamma$ and the L points.
Overall, we can say that 
the discontinuous and the split branches occur because
different combinations of the chemical elements contribute to different regions of frequency.


In Fig.~\ref{fig:phonons_Cu075Au025},
we also compare the phonon band structure of Cu$_{0.75}$Au$_{0.25}$ 
calculated using the band-unfolding method 
with those calculated using the itinerant coherent potential approximation (ICPA) method
\cite{Ghosh2002}.
In the ICPA method,
the phonon band structure of a disordered system is calculated with consideration in 
the variations of both atomic masses and force constants among the atomic sites,
like the band-unfolding method,
while the ICPA method is based not on the supercell approach 
but on the augmented-space formalism
\cite{Mookerjee1973}.
%
The peak positions of the spectral function of the band-unfolding method are 
mostly in agreement with those of the ICPA method.
The result of the ICPA method shows the discontinuity in the lowest-frequency branch along the $\langle 110 \rangle$ direction as well as the result of the band-unfolding method.



Finally, in Fig.~\ref{fig:phonons_Cu075Au025},
the phonon band structure of Cu$_{0.75}$Au$_{0.25}$
calculated using the band-unfolding method is mostly in agreement with experimental data
\cite{Katano1988}.
The experimental data, however, do not clearly show
the discontinuous branch along the $\langle 110 \rangle$ direction or the split branch along the $\langle 111 \rangle$ direction unlike the result of the band-unfolding method.
The reason of this discrepancy is not clear,
but since both the band-unfolding and the ICPA methods show the discontinuous
branch along the $\langle 110 \rangle$ direction,
we think that
the discontinuous branch can be found in computational approaches
as far as we incorporate the variations of atomic masses and force constants 
among the atomic sites into the calculations.
It should be noted that
discontinuous phonon branches have been observed also in experiments
for disordered fcc Ni--Pt \cite{Tsunoda1979} and Cu--Pt \cite{Kunitomi1980} alloys,
which have large variations of atomic masses and, possibly, force constants
among the atomic sites
as well as disordered fcc Cu$_{0.75}$Au$_{0.25}$.
This fact implies that
the experimental data for Cu$_{0.75}$Au$_{0.25}$
might have overlooked the peculiar behaviors found in the computational approaches.
It should be also mentioned that
the effective phonon band structure of Cu$_{0.25}$Au$_{0.25}$ calculated 
using the average atomic mass and force constants (or dynamical matrices)
over the chemical elements
\cite{Wang2011}
seems to be in good agreement with the experimental data.
However,
the approach in Ref.~\cite{Wang2011} ignores
the variations of atomic masses and force constants among the atomic sites
inherent in disordered alloys 
and hence does not sufficiently describe the actual situation
in disordered alloys.
As shown above,
these variations are large in Cu$_{0.75}$Au$_{0.25}$
and hence should be explicitly incorporated into calculations of 
the effective phonon band structure.
The computational result in Ref.~\cite{Wang2011} using the average values 
may be accidentally in agreement with the experimental data.


\section{CONCLUSIONS}
\label{sec:conclusions}

In this study,
we develop a procedure to decompose the effective band structures
obtained using the band-unfolding method
according to the SRs of the little groups.
For the decomposition,
we derive the projection operators for SRs based on a group-theoretical approach.
The current procedure enables us to compare the band structure of 
a disordered system with that of an ordered system or of another disordered system 
in a consistent manner in terms of crystallographic symmetry.
We also derive the projection operators for chemical elements,
which enables us to investigate the contributions of different combinations 
of chemical elements to the unfolded band structures.

We apply the current band-unfolding method to the phonon band structure of disordered fcc Cu$_{0.75}$Au$_{0.25}$,
which has large variations of atomic masses and force constants among the atomic sites
due to the chemical disorder.
The calculated phonon band structure shows the linewidths of phonon modes induced by the chemical disorder in Cu$_{0.75}$Au$_{0.25}$.
The phonon band structure also shows several peculiar behaviors such as the
discontinuous and the split branches for the modes of specific SRs.
These peculiar behaviors occur because different combinations 
of the chemical elements contribute to different regions of frequency for these branches.

The band-unfolding method can be applied not only to systems with chemical disorder 
but in principle also to those with magnetic disorder.
Recently several computational approaches have been attempted 
to obtain the phonon band structures of magnetic systems 
in the high-temperature paramagnetic (PM) phase
\cite{
Kormann2012, Kormann2014, Ikeda2014,  
Kormann2016Impact,  
Shulumba2014, Zhou2014,  
Leonov2012, Leonov2014Electronic}.  
It is possible to also employ the band-unfolding method 
for obtaining the phonon band structures in the PM phase 
modeled by a supercell model with disordered magnetic moments,
which may enable us to estimate 
the impact of thermal magnetic fluctuation on the phonon band structures.


\begin{acknowledgments}

Funding by the Ministry of Education, Culture, Sports,
Science and Technology (MEXT), Japan, through Elements Strategy Initiative for
Structural Materials (ESISM) of Kyoto University,
and by the Japan Society for the Promotion of Science (JSPS) KAKENHI 
Grant-in-Aid for Young Scientist (B) (Grant No. 16K18228)
are gratefully acknowledged.

\end{acknowledgments}

\appendix
\section{Transformation of functions}
\label{sec:transformation_of_functions}
  
When 
$\{ \mathbf{R} | \mathbf{w} \}$
is applied to a scalar-field function $f (\mathbf{x})$,
the transformed function
$f' (\mathbf{x}) \equiv \{ \mathbf{R} | \mathbf{w} \} f (\mathbf{x})$
satisfies
\begin{align}
f'(\{ \mathbf{R} | \mathbf{w} \}\mathbf{x})
&=
f(\mathbf{x}).
\end{align}
Therefore,
\begin{align}
f'(\mathbf{x})
&=
f(\{ \mathbf{R} | \mathbf{w} \}^{-1} \mathbf{x})
\notag
\\
&=
f (\mathbf{R}^{-1} \mathbf{x} - \mathbf{R}^{-1} \mathbf{w})
,
\label{eq:tf_sff}
\end{align}
where 
$
\{ \mathbf{R} | \mathbf{w} \}^{-1}
\equiv
\{ \mathbf{R}^{-1} | - \mathbf{R}^{-1} \mathbf{w} \}
$
is the inverse transformation operator to
$ \{ \mathbf{R} | \mathbf{w} \} $.
Similarly, when
$\{ \mathbf{R} | \mathbf{w} \}$
is applied to a vector-field function $\mathbf{f} (\mathbf{x})$,
the transformed function
$\mathbf{f}' (\mathbf{x}) \equiv \{ \mathbf{R} | \mathbf{w} \} \mathbf{f} (\mathbf{x})$
satisfies
\begin{align}
\mathbf{f}'
(\{ \mathbf{R} | \mathbf{w} \}\mathbf{x})
&=
\mathbf{R} \mathbf{f}
(\mathbf{x}).
\end{align}
Therefore,
\begin{align}
\mathbf{f}'(\mathbf{x})
&=
\mathbf{R} \mathbf{f}
(\{ \mathbf{R} | \mathbf{w} \}^{-1} \mathbf{x})
\notag
\\
&=
\mathbf{R} \mathbf{f}
(\mathbf{R}^{-1} \mathbf{x} - \mathbf{R}^{-1} \mathbf{w})
.
\label{eq:tf_vff}
\end{align}

\section{Orthogonality relations for SRs}
\label{sec:orthogonality_relations_for_SRs}

Here we show the orthogonality relations for the SRs of the little group 
${\mathcal{G}}^{\mathbf{k}}$ of the wave vector $\mathbf{k}$,
which are used to derive the projection operators for SRs
in Eq.~\eqref{eq:po_srs}.
Since ${\mathcal{G}}^{\mathbf{k}}$ is a infinite group,
we cannot directly use the orthogonality relations for IRs for finite groups.
As shown below, however,
we can derive the orthogonality relations for the coset representatives 
of ${\mathcal{G}}^{\mathbf{k}}$ relative to the translation subgroup
$\mathcal{T}$,
which is very similar to the orthogonality relations for finite groups.
First we show that the SRs of $\mathcal{G}^{\mathbf{k}}$ can be written 
using the IPRs of the little cogroup $\lcg$
in the same manner as in the literature
(e.g., Sec.~14.4.2 in Ref.~\cite{Kim1999}).
Then we derive the orthogonality relations for the SRs using 
the orthogonality relations for the IPRs of $\lcg$.

$\mathcal{G}^{\mathbf{k}}$ is decomposed using the coset representatives 
$\{ \mathbf{R}_{j} | \mathbf{w}_{j} \}$ relative to $\mathcal{T}$,
as shown in Eq.~\eqref{eq:coset_representatives_Gk}.
The coset representatives satisfy the following multiplication rule;
\begin{align}
\{ \mathbf{R}_{j} | \mathbf{w}_{j} \}
\{ \mathbf{R}_{k} | \mathbf{w}_{k} \}
&=
\{ \mathbf{I}_{3} | \mathbf{t}     \}
\{ \mathbf{R}_{l} | \mathbf{w}_{l} \},
\end{align}
where $\mathbf{R}_{l} = \mathbf{R}_{j} \mathbf{R}_{k}$
and
$
    \mathbf{t}
    =
    \mathbf{R}_{j} \mathbf{w}_{k} +
    \mathbf{w}_{j} -
    \mathbf{w}_{l}
    \in
    \mathcal{T}.
$
The $\mu$th SR $\Gamma^{\mathbf{k}\mu}$ of 
$\mathcal{G}^{\mathbf{k}}$ then satisfies
\begin{align}
    \Gamma^{\mathbf{k}\mu} (\{ \mathbf{R}_{j} | \mathbf{w}_{j} \})
    \Gamma^{\mathbf{k}\mu} (\{ \mathbf{R}_{k} | \mathbf{w}_{k} \})
    &=
    e^{-i \mathbf{k} \cdot \mathbf{t}}
    \Gamma^{\mathbf{k}\mu} (\{ \mathbf{R}_{l} | \mathbf{w}_{l} \}),
    \label{eq:mult_small_rep}
\end{align}
where the property of SRs in Eq.~\eqref{eq:def_srs} is applied to
$\{ \mathbf{I}_{3} | \mathbf{t} \}$.
Suppose that
$\Gamma^{\mathbf{k}\mu} (\{ \mathbf{R} | \mathbf{w} \})$
$(\{ \mathbf{R} | \mathbf{w} \} \in \mathcal{G}^{\mathbf{k}})$ 
is decomposed as
\begin{align}
    \Gamma^{\mathbf{k} \mu}
    (\{ \mathbf{R} | \mathbf{w} \})
    &=
    e^{-i \mathbf{k} \cdot \mathbf{w}}
    {\Delta}^{\mathbf{k} \mu}
    (\mathbf{R})
    .
    \label{eq:small_rep_and_proj_rep}
\end{align}
By substituting Eq.~(\ref{eq:small_rep_and_proj_rep}) into 
Eq.~(\ref{eq:mult_small_rep}),
\begin{gather}
    e^{-i \mathbf{k} \cdot \mathbf{w}_{j}}
    \Delta^{\mathbf{k}\mu} (\mathbf{R}_{j})
    e^{-i \mathbf{k} \cdot \mathbf{w}_{k}}
    \Delta^{\mathbf{k}\mu} (\mathbf{R}_{k})
    =
    e^{-i \mathbf{k} \cdot \mathbf{t}}
    e^{-i \mathbf{k} \cdot \mathbf{w}_{l}}
    \Delta^{\mathbf{k}\mu} (\mathbf{R}_{l})
    .
    \\
    \therefore
    \,
    \Delta^{\mathbf{k} \mu} (\mathbf{R}_{j})
    \Delta^{\mathbf{k} \mu} (\mathbf{R}_{k})
    =
    \lambda (j, k)
    \Delta^{\mathbf{k} \mu} (\mathbf{R}_{j} \mathbf{R}_{k})
    ,
    \label{eq:mult_proj_rep}
\end{gather}
where
$
    \lambda (j, k)
    \equiv
    e^{-i (\mathbf{R}^{T}_{j} \mathbf{k} - \mathbf{k}) \cdot \mathbf{w}_{k}}
$.
$\lambda (j, k)$ is found to satisfy
$\lambda (j, k) \lambda (jk, m) = \lambda (j, km) \lambda (k, m)$.
Eq.~(\ref{eq:mult_proj_rep}) therefore indicates that
$\Delta^{\mathbf{k}\mu}$ is a projective representation (PR)
\cite{ITD_2013, Kim1999}
of the little cogroup $\lcg$ with the factor system defined as the set of $\lambda (j, k)$.
$\Delta^{\mathbf{k}\mu}$ is irreducible 
because $\Gamma^{\mathbf{k}\mu}$ is supposed to be irreducible.
Now $|\lambda (j, k)|$ is equal to one for all the combinations of the coset representatives,
and hence the IPR $\Delta^{\mathbf{k}\mu}$ can be transformed to be unitary 
\textit{without changing the factor system}.
This can be proved in a similar manner to that in, e.g., Theorem~12.3.1 in Ref.~\cite{Kim1999}.
$\Delta^{\mathbf{k}\mu}$ can therefore be supposed to be unitary without loss of generality.
Note that in special cases
the set of the coset representatives can be chosen
so that $\lambda(j, k)$
is equal to one for all the combinations of the coset representatives.
In such cases 
$\Delta^{\mathbf{k} \mu}$
reduces to an ordinary IR of $\lcg$.
Such cases occur, e.g.,
when the space group $\mathcal{G}$ is symmorphic
or when $\mathbf{k}$ is not on the BZ boundary.

The set of the unitary IPRs $\{ \Delta^{\mathbf{k}\mu} \}$ 
belonging to the same factor system satisfies the following orthogonality relations
(see, e.g., Theorem~12.3.2 in Ref.~\cite{Kim1999});
\begin{align}
    \sum_{\mathbf{R} \in \bar{\mathcal{G}^{\mathbf{k}}}}
    \Delta^{\mathbf{k} \mu}_{j k } (\mathbf{R}) 
    \Delta^{\mathbf{k} \nu}_{j'k'} (\mathbf{R})^{*}
    &=
    \frac
    {|\bar{\mathcal{G}}^{\mathbf{k}}|}
    {d_{\mu}}
    \delta_{\mu\nu}
    \delta_{j j'}
    \delta_{k k'}
    .
    \label{eq:orthogonality_little_cogroup}
\end{align}
Using Eq.~(\ref{eq:orthogonality_little_cogroup})
it is also found that
the set of the SRs $\{ \Gamma^{\mathbf{k}\mu} \}$ satisfies 
the following orthogonality relations for the coset representatives;
\begin{align}
    &
    \sum_{l = 1}^{n}
    \Gamma^{\mathbf{k} \mu}_{j k } (\{ \mathbf{R}_{l} | \mathbf{w}_{l} \}) 
    \Gamma^{\mathbf{k} \nu}_{j'k'} (\{ \mathbf{R}_{l} | \mathbf{w}_{l} \})^{*}
    \notag
    \\
    &\quad
    =
    \sum_{l = 1}^{n}
    e^{-i \mathbf{k} \cdot \mathbf{w}_{l}} \Delta^{\mathbf{k} \mu}_{j k } (\mathbf{R}_{l}) 
    e^{ i \mathbf{k} \cdot \mathbf{w}_{l}} \Delta^{\mathbf{k} \nu}_{j'k'} (\mathbf{R}_{l})^{*}
    \notag
    \\
    &\quad=
    \frac
    {|\bar{\mathcal{G}}^{\mathbf{k}}|}
    {d_{\mu}}
    \delta_{\mu\nu}
    \delta_{j j'}
    \delta_{k k'}
    .
    \label{eq:orthogonality_sr}
\end{align}
It should be emphasized that although Eq.~\eqref{eq:orthogonality_sr}
looks similar to Eq.~\eqref{eq:orthogonality_little_cogroup},
Eq.~\eqref{eq:orthogonality_sr} is for the little group $\mathcal{G}^{\mathbf{k}}$,
which is an infinite group.

\section{Derivation of Eq.~(\ref{eq:effect_of_pos_wvs_simple})}
\label{sec:effect_of_pos_wvs}

The notations follow those in Sec.~\ref{sec:pos_wvs}.
\begin{align}
    \hat{P}^{\mathbf{k}_{k}}
    f^{\mathbf{K}} (\mathbf{x})
    &=
    \frac{1}{N}
    \sum_{j = 1}^{N}
    e^{i \mathbf{k}_{k} \cdot \mathbf{t}_{j}}
    \hat{t}_{j}
    \sum_{l = 1}^{N}
    c_{\mathbf{k}_{l}}
    f^{\mathbf{k}_{l}} (\mathbf{x})
    \notag
    \\
    &=
    \frac{1}{N}
    \sum_{j = 1}^{N}
    e^{i \mathbf{k}_{k} \cdot \mathbf{t}_{j}}
    \sum_{l = 1}^{N}
    c_{\mathbf{k}_{l}}
    \left[
    \hat{t}_{j}
    f^{\mathbf{k}_{l}} (\mathbf{x})
    \right]
    \notag
    \\
    &=
    \frac{1}{N}
    \sum_{j = 1}^{N}
    e^{i \mathbf{k}_{k} \cdot \mathbf{t}_{j}}
    \sum_{l = 1}^{N}
    c_{\mathbf{k}_{l}}
    f^{\mathbf{k}_{l}} (\mathbf{x})
    e^{- i \mathbf{k}_{l} \cdot \mathbf{t}_{j}}
    \,\, [\because \textrm{Eq.~\eqref{eq:transformation_wv}}]
    \notag
    \\
    &=
    \sum_{l = 1}^{N}
    c_{\mathbf{k}_{l}}
    f^{\mathbf{k}_{l}} (\mathbf{x})
    \left[
    \frac{1}{N}
    \sum_{j = 1}^{N}
    e^{i (\mathbf{k}_{k} - \mathbf{k}_{l}) \cdot \mathbf{t}_{j}}
    \right]
    \notag
    \\
    &=
    \sum_{l = 1}^{N}
    c_{\mathbf{k}_{l}} f^{\mathbf{k}_{l}} (\mathbf{x}) \delta_{kl}
    \,\,
    [\because \textrm{Eq.}~\eqref{eq:fourier_k}]
    \notag
    \\
    &=
    c_{\mathbf{k}_{k}} f^{\mathbf{k}_{k}} (\mathbf{x})
    .
    \label{eq:effect_of_pos_wvs}
\end{align}

\section{Derivation of Eq.~(\ref{eq:effect_of_pos_SRs_simple})}
\label{sec:effect_of_projection_operators_sr}

The notations follow those in Sec.~\ref{sec:pos_srs}.
\begin{align}
    \hat{P}^{\mathbf{k}\mu}
    f^{\mathbf{k}} (\mathbf{x})
    &=
    \frac{d_{\mu}}{| \lcg |}
    \sum_{j = 1}^{| \lcg |}
    \chi^{\mathbf{k}\mu} (\hat{g}_{j})^{*}
    \hat{g}_{j}
    \sum_{\nu}
    \sum_{s = 1}^{n_{\nu}}
    \sum_{k = 1}^{d_{\nu}}
    c_{\nu s k}
    f^{\mathbf{k} \nu s k}
    (\mathbf{x})
    \notag
    \\
    &=
    \frac{d_{\mu}}{| \lcg |}
    \sum_{j = 1}^{| \lcg |}
    \chi^{\mathbf{k}\mu} (\hat{g}_{j})^{*}
    \sum_{\nu}
    \sum_{s = 1}^{n_{\nu}}
    \sum_{k = 1}^{d_{\nu}}
    c_{\nu s k}
    \left[
    \hat{g}_{j}
    f^{\mathbf{k} \nu s k}
    (\mathbf{x})
    \right]
    \notag
    \\
    &=
    \frac{d_{\mu}}{| \lcg |}
    \sum_{j = 1}^{| \lcg |}
    \chi^{\mathbf{k}\mu} (\hat{g}_{j})^{*}
    \sum_{\nu}
    \sum_{s = 1}^{n_{\nu}}
    \sum_{k = 1}^{d_{\nu}}
    c_{\nu s k}
    \notag
    \\
    &\times
    \left[
    \sum_{l = 1}^{d_{\nu}}
    f^{\mathbf{k} \nu s l}
    (\mathbf{x})
    \Gamma_{lk}^{\nu}
    (\hat{g}_{j})
    \right]
    \quad
    [\because \textrm{Eq.~(\ref{eq:transformation_sr})}]
    \notag
    \\
    &=
    \sum_{m = 1}^{d_{\mu}}
    \sum_{\nu}
    \sum_{s = 1}^{n_{\nu}}
    \sum_{k = 1}^{d_{\nu}}
    c_{\nu s k}
    \sum_{l = 1}^{d_{\nu}}
    f^{\mathbf{k} \nu s l}
    (\mathbf{x})
    \notag
    \\
    &\times
    \left[
    \frac{d_{\mu}}{| \lcg |}
    \sum_{j = 1}^{| \lcg |}
    \Gamma_{mm}^{\mu}
    (\hat{g}_{j})^{*}
    \Gamma_{lk}^{\nu}
    (\hat{g}_{j})
    \right]
    \notag
    \\
    &=
    \sum_{m = 1}^{d_{\mu}}
    \sum_{\nu}
    \sum_{s = 1}^{n_{\nu}}
    \sum_{k = 1}^{d_{\nu}}
    c_{\nu s k}
    \sum_{l = 1}^{d_{\nu}}
    f^{\mathbf{k} \nu s l}
    (\mathbf{x})
    \delta_{\mu \nu}
    \delta_{m l}
    \delta_{m k}
    \notag
    \\
    &\quad
    [\because \textrm{Eq.~(\ref{eq:orthogonality_sr})}]
    \notag
    \\
    &=
    \sum_{s = 1}^{n_{\mu}}
    \sum_{k = 1}^{d_{\mu}}
    c_{\mu s k}
    f^{\mathbf{k} \mu s k}
    (\mathbf{x})
    .
\end{align}

\section{Unfolded phonon band structure calculated using empirical
interatomic potentials}
\label{sec:empirical_potentials}

Here we investigate the convergence of the unfolded phonon band structure of disordered fcc Cu$_{0.75}$Au$_{0.25}$
with respect to the supercell size using the empirical embedded-atom-method (EAM) interatomic potentials.
In Sec.~\ref{sec:results},
the unfolded phonon band structure of disordered fcc Cu$_{0.75}$Au$_{0.25}$ is obtained from first-principles calculations using the $2 \times 2 \times 2$ and the $3 \times 3 \times 3$ supercell models.
Although we can find several peculiar behaviors such as the discontinuous and the split branches 
in the unfolded phonon band structure,
one may wonder if these peculiar behaviors are spurious due to the limited supercell size.
Therefore it is worth confirming the cell-size convergence of the unfolded phonon band structure.
It, however, requires prohibitively high computational costs to calculate the second-order force constants for further larger supercell models of disordered fcc Cu$_{0.75}$Au$_{0.25}$ based on first-principles.
The use of the EAM interatomic potentials enables us to access the further larger supercell models 
because it requires much less computational costs to calculate the second-order force constants than first-principles calculations.

Disordered fcc Cu$_{0.75}$Au$_{0.25}$ was modeled using the
$2 \times 2 \times 2$,
$3 \times 3 \times 3$,
$4 \times 4 \times 4$,
$5 \times 5 \times 5$, and
$6 \times 6 \times 6$
supercells of the conventional fcc unit cell.
For the $2 \times 2 \times 2$ and the $3 \times 3 \times 3$ supercell models,
the chemical disorder was approximated using the SQSs the same as those used in the first-principles calculations,
while for the further larger supercell models,
the chemical disorder was approximated using a pseudorandom-number generator.
We used the EAM interatomic potentials parametrized by
\citeauthor{Foiles1986}~\cite{Foiles1986}
as implemented in the \textsc{LAMMPS} code
\cite{Plimpton1995}.
The lattice shape was kept cubic,
and the lattice constant of Cu$_{0.75}$Au$_{0.25}$ was fixed to the
experimental value at room temperature,
3.753~{\AA}~\cite{Pearson1958}.
The internal atomic positions were optimized
until the residual forces became less than
$1 \times 10^{-9}$~eV/{\AA}.
The second-order force constants of the supercell models were calculated 
using finite atomic displacements of 0.01~{\AA} with no further expansion of
the supercell models.
Phonon modes obtained from the supercell models were unfolded into the BZ 
for the primitive fcc unit cell.

\begin{figure}[tbp]
\centering
\includegraphics[width=\linewidth]{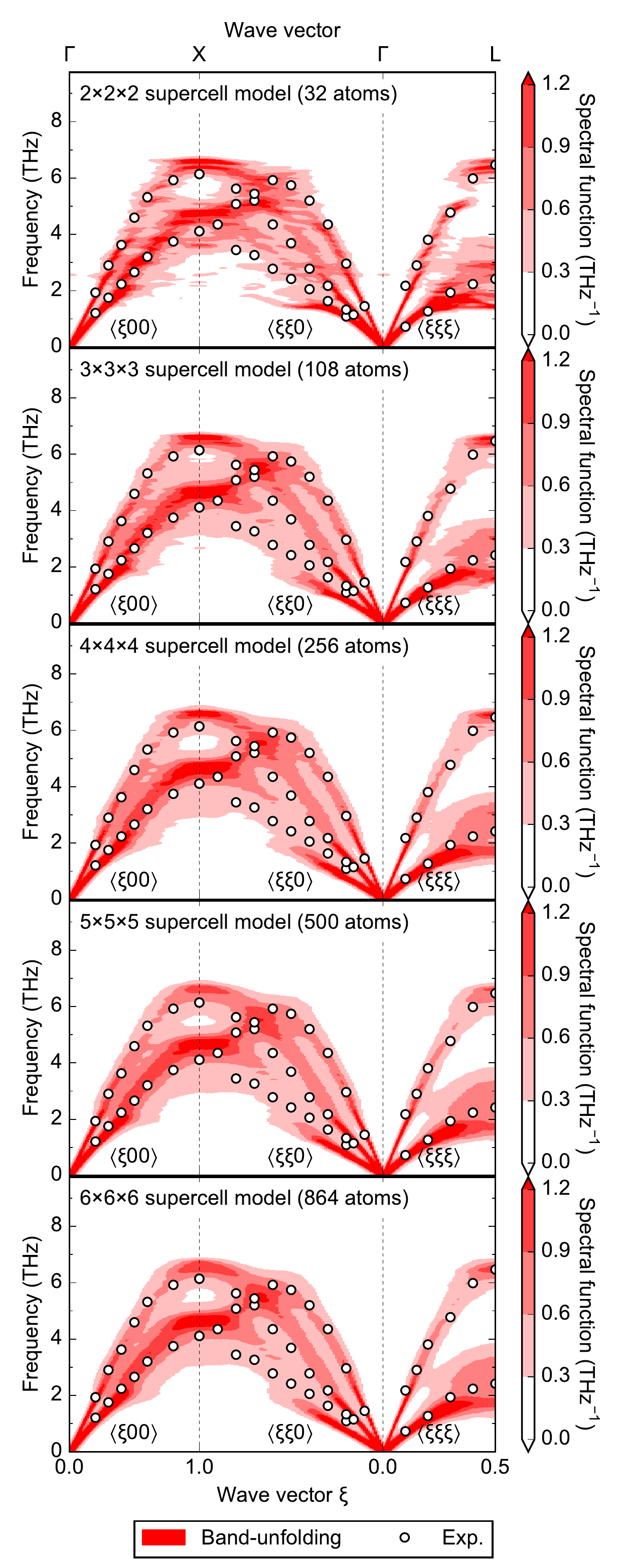}
\caption{
(Color online)
Unfolded phonon band structure of disordered fcc Cu$_{0.75}$Au$_{0.25}$
calculated using the EAM interatomic potentials.
The sizes of the supercell models are shown at the upper-left in the panels.
White circles represent experimental data at room temperature
\cite{Katano1988}.
\label{fig:phonons_lammps_Cu075Au025}
}
\end{figure}
%
Figure~\ref{fig:phonons_lammps_Cu075Au025} shows the unfolded phonon band structure
of disordered fcc Cu$_{0.75}$Au$_{0.25}$ calculated using the EAM interatomic potentials.
The results are qualitatively very similar to those obtained 
using first-principles calculations shown in Fig.~\ref{fig:phonons_Cu075Au025}.
%
%
The spectral function is almost converged at the $3 \times 3 \times 3$ supercell model;
the result of the $3 \times 3 \times 3$ supercell model is very similar to 
the result of the $6 \times 6 \times 6$ supercell model.
Actually,
even the result of the $2 \times 2 \times 2$ supercell model captures 
most characteristics of the spectral functions of the larger supercell models.
The discontinuous branch along the $\langle 110 \rangle$ direction and 
the split branch along the $\langle 111 \rangle$ direction
are found in the EAM results,
as well as the first-principles results, 
even for the $6 \times 6 \times 6$ supercell model.
Since the spectral function is expected to be converged at
the $6 \times 6 \times 6$ supercell model,
this result implies that 
the discontinuous and the split branches are not spurious behaviors 
due to the limited supercell size 
but realistic ones originating from the chemical disorder in Cu$_{0.75}$Au$_{0.25}$.
As analyzed in Sec.~\ref{sec:results},
the discontinuous and the split branches occur
because different combinations of the chemical elements contribute to 
different regions of frequency for specific modes.


\section{Variations of force constants in Cu$_{0.75}$Au$_{0.25}$}
\label{sec:force_constants}

\begin{figure}[tbp]
\centering
\includegraphics[width=\linewidth]{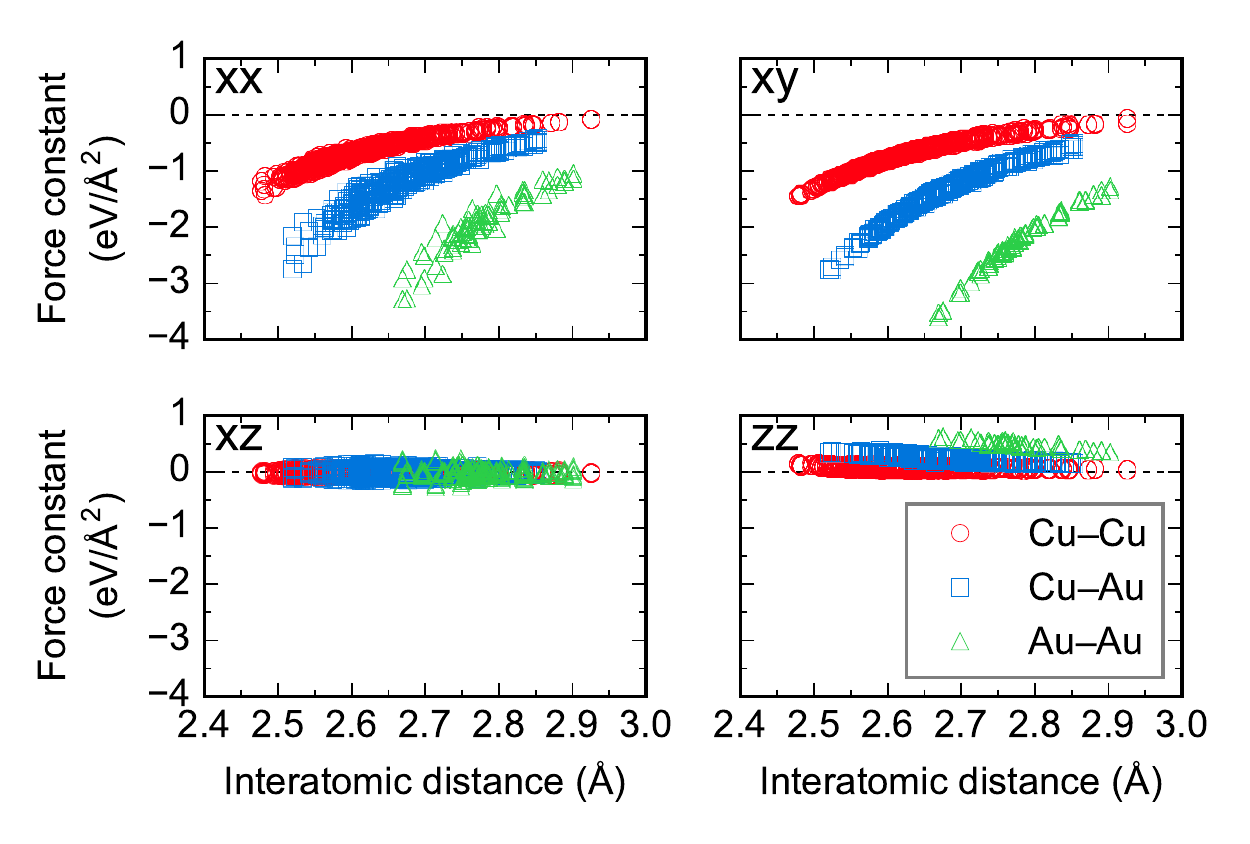}
\caption{
(Color online)
Distributions of the second-order force constants between the 1NN atomic pairs
with respect to interatomic distance
for disordered fcc Cu$_{0.75}$Au$_{0.25}$ obtained from 
the $3 \times 3 \times 3$ supercell model.
Red circles, blue squares, and green triangles are
for Cu--Cu, Cu--Au (Au--Cu), and Au--Au pairs, respectively.
The first and the second chemical components are supposed to be
on $(0, 0, 0)$ and $(1/2, 1/2, 0)$, 
respectively,
in fractional coordinates for the conventional fcc unit cell.
Each panel corresponds to the symmetrically-inequivalent element 
of the force constants 
specified at the upper-left in the panel in Cartesian coordinates,
where the first and the second symbols are 
for the first and the second chemical components, respectively.
\label{fig:fc_Cu075Au025}
}
\end{figure}

\begin{table*}[tb]
\centering
\caption{
Average and standard deviation (SD) of the second-order force constants 
and interatomic distances between the 1NN atomic pairs.
\label{tb:fc_Cu075Au025}
}
\begin{ruledtabular}
\begin{tabular}{cdddddddddd}
       & 
\multicolumn{2}{c}{Interatomic distance (\AA)} &
\multicolumn{8}{c}{Force constants (eV/\AA$^{2}$)} \\
 & & &
\multicolumn{2}{c}{xx} &
\multicolumn{2}{c}{xy} &
\multicolumn{2}{c}{xz} &
\multicolumn{2}{c}{zz} \\
&
\multicolumn{1}{c}{Average} &
\multicolumn{1}{c}{SD} &
\multicolumn{1}{c}{Average} &
\multicolumn{1}{c}{SD} &
\multicolumn{1}{c}{Average} &
\multicolumn{1}{c}{SD} &
\multicolumn{1}{c}{Average} &
\multicolumn{1}{c}{SD} &
\multicolumn{1}{c}{Average} &
\multicolumn{1}{c}{SD} \\
\hline
Cu--Cu & 2.626 & 0.078 & -0.636 & 0.237 & -0.718 & 0.261 & 0.000 & 0.028 & 0.086 & 0.020 \\
Cu--Au & 2.679 & 0.070 & -1.165 & 0.405 & -1.368 & 0.441 & 0.000 & 0.052 & 0.256 & 0.048 \\
Au--Au & 2.776 & 0.055 & -1.919 & 0.497 & -2.279 & 0.532 & 0.000 & 0.084 & 0.502 & 0.066 \\
\end{tabular}
\end{ruledtabular}
\end{table*}

Figure~\ref{fig:fc_Cu075Au025} shows the distributions of the second-order 
force constants between the first nearest-neighbor (1NN) atomic pairs
with respect to interatomic distance
for disordered fcc Cu$_{0.75}$Au$_{0.25}$
calculated from the $3 \times 3 \times 3$ supercell model.
Table~\ref{tb:fc_Cu075Au025} summarizes the average and the standard deviation
of the force constants.

The force constants clearly depend on the combinations of the chemical elements.
At a certain interatomic distance,
the force constants of the Cu--Cu pairs are smaller in magnitude than
those of the Cu--Au (Au--Cu) and the Au--Au pairs.
The strong dependence of the force constants on the combinations of the
chemical elements indicates that
it is not adequate to take the average of force constants over the combinations of the chemical elements 
to describe the real physics in Cu$_{0.75}$Au$_{0.25}$.


The force constants also depend on the interatomic distance.
The force constants tend to be smaller in magnitude as the bond distance increases.

%

\end{document}